\documentclass[aps,pre,twocolumn,groupedaddress,longbibliography,floats,final,postprint,superscriptaddress]{revtex4-2}

\usepackage{xcolor}
\usepackage[export]{adjustbox}
\usepackage{graphicx}
\usepackage{amsmath}
\usepackage{amssymb}
\usepackage{amsthm}
\usepackage{bm}
\usepackage[citecolor=blue,linkcolor=blue,colorlinks]{hyperref}
\usepackage{hypcap} % fix the links
\usepackage{threeparttable}
\usepackage{enumerate}
\usepackage{physics}
\makeatletter
\let\old@makecaption=\@makecaption
\usepackage{subcaption}
\let\@makecaption=\old@makecaption
\makeatother
\usepackage{float}
\usepackage{multirow}
\usepackage{booktabs}
\usepackage{mathtools}
\usepackage{textcomp}
\usepackage{cleveref}

\definecolor{blue}{rgb}{0,0,1}
\definecolor{red}{rgb}{1,0,0}
\definecolor{green}{rgb}{0,1,0}
\definecolor{darkgreen}{rgb}{0,0.4,0}
\definecolor{darkred}{rgb}{0.4,0,0}
\definecolor{darkblue}{rgb}{0,0,0.4}

\newcommand{\scrA}{{\mathcal A}}

\newcommand{\scrG}{{\mathcal G}}
\newcommand{\scrI}{{\mathcal I}}
\newcommand{\scrL}{{\mathcal L}}
\newcommand{\scrM}{{\mathcal M}}

\newcommand{\scrS}{{\mathcal S}}

\newcommand{\scrT}{{\mathcal T}}

\newcommand{\scrZ}{{\mathcal Z}}

\newcommand{\bco}{\beta_{\rm c1}}
\newcommand{\bct}{\beta_{\rm c2}}
\newcommand{\chih}{\chi_{\textsc{h}}}
\newcommand{\chil}{\chi_{\textsc{l}}}
\newcommand{\RNum}[1]{\expandafter{\romannumeral #1\relax}}

\graphicspath{{figures/}}
\begin{document}

\title{Monte Carlo study of duality and the Berezinskii-Kosterlitz-Thouless phase transitions of the two-dimensional q-state clock model in flow representations}
\author{Hao Chen}
\thanks{These two authors contributed equally to this paper.}
\affiliation{School of the Gifted Young, University of Science and Technology of China, Hefei, Anhui 230026, China}
\author{Pengcheng Hou}
\thanks{These two authors contributed equally to this paper.}
\affiliation{Department of Modern Physics, University of Science and Technology of China, Hefei, Anhui 230026, China}	
\author{Sheng Fang}
\email{fs4008@mail.ustc.edu.cn}
\affiliation{Department of Modern Physics, University of Science and Technology of China, Hefei, Anhui 230026, China}	
\author{Youjin Deng}
\email{yjdeng@ustc.edu.cn}
\affiliation{Department of Modern Physics, University of Science and Technology of China, Hefei, Anhui 230026, China}	
\affiliation{
Shanghai Research Center for Quantum Sciences, Shanghai 201315, China}
\affiliation{MinJiang Collaborative Center for Theoretical Physics,
	College of Physics and Electronic Information Engineering, Minjiang University, Fuzhou 350108, China}

\begin{abstract}
	
The two-dimensional $q$-state clock model for $q \geq 5$ undergoes two Berezinskii-Kosterlitz-Thouless (BKT) phase transitions as temperature decreases. Here we report an extensive worm-type simulation of the square-lattice clock model for $q=$5--9 in a pair of flow representations, from the high- and low-temperature expansions, respectively. By finite-size scaling analysis of  susceptibility-like quantities, 
we determine the critical points with a precision improving over the existing results. 
Due to the dual flow representations, each point in the critical region is observed to simultaneously exhibit a pair of anomalous dimensions, which are $\eta_1=1/4$ and $\eta_2 = 4/q^2$ at the two BKT transitions. 
Further, the approximate self-dual points $\beta_{\rm sd}(L)$, defined by the stringent condition that the susceptibility like quantities in both flow representations are identical, 
are found to be nearly independent of system size $L$ and behave as $\beta_{\rm sd} \simeq q/2\pi$ asymptotically at the large-$q$ limit.
The exponent $\eta$ at $\beta_{\rm sd}$ is consistent with $1/q$ within statistical error 
as long as $q \geq 5$. Based on this, we further conjecture that $\eta(\beta_{\rm sd}) = 1/q$ holds exactly and is universal for systems in the $q$-state clock universality class. Our work provides a vivid demonstration of rich phenomena associated with the duality and self-duality of the clock model in two dimensions.
\end{abstract}
\maketitle

\section{Introduction}
\label{Sec:Introduction}
The $q$-state clock model is a prototypical model in the study of phase transitions due to its rich critical phenomena.
It can be seen as a discretized version of the $XY$ model, where the classical spin ${ \bf{S} } = (\cos\theta, \sin \theta)$ on each site is confined in a two-dimensional plane and takes one of the $q$ 
uniform orientations specified by the angle $\theta = 2\pi\sigma /q $ with integer $\sigma\in \{0,1, \dots, q-1\}$. 
Neighboring spins in the model are coupled via the form $-J {\bf {S}}_i \cdot {\bf{S}}_j$
and the partition function is written as 
    \begin{equation}
        \label{eq:partition_function}
        \scrZ = \sum_{ \{ \bf S \} } \prod_{\langle  ij \rangle } e^{\beta J {\bf{S}}_i \cdot {\bf{S}}_j  } = \sum_{\{ \sigma \}} \prod_{\langle  ij \rangle } e^{\beta J \cos[2\pi/q (\sigma_i - \sigma_j) ]}\;,
    \end{equation}
where the summation is over all possible spin configurations, $\beta$ is the inverse temperature, $J$ is the coupling strength, and $\langle  ij \rangle$ stands for neighboring pairs. 

\begin{figure}[h]
	\centering
	\includegraphics[width=1.\columnwidth]{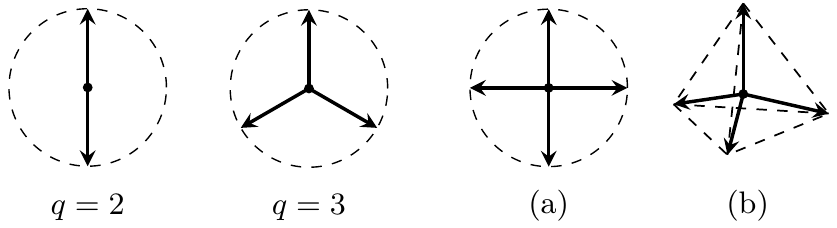}
	\caption{
	Unit-vectors of the $q$-state clock model and the Potts model. For $q=2$ and 3, the possible directions
	taken by spins are identical in both models. They become different for $q\geq4$, and (a) and (b) correspond to the 
	four-state clock model and the four-state Potts model, respectively.}
	\label{fig:Potts}
\end{figure}

The $q$-state clock model possesses a discrete $\mathbb{Z}_q$ symmetry and recovers to the $XY$ model with $\text{U}(1)$ symmetry in the $q\! \to \! \infty$ limit. 
It is worth mentioning that the $q$-state clock model is also referred to as the planar Potts model or the vector Potts model \cite{wuPottsModel1982} due to 
its similarity to the standard Potts model.
In the standard Potts model, spins also take $q$ different values, but the neighboring spins are coupled as $- J_\textsc{p} \delta_{\sigma_i, \sigma_j}$. Hence, the standard Potts model has $\mathbb{S}_q$ symmetry instead of     $\mathbb{Z}_q$ symmetry.
The clock model and the Potts model are identical for $q=2$ and $q=3$ with $J = \frac{1}{2} J_\textsc{p}$ and $J = \frac{2}{3} J_\textsc{p}$, respectively.
However, the two models can no longer be mapped to each other due to their intrinsic symmetries for $q \geq 4$. For example, the four-state clock model can be 
mapped onto two decoupled Ising models, different from the four-state Potts model, as shown in Fig.~\ref{fig:Potts}. 
For convenience, we will take the coupling strength  $J=1$ in the following. 
\par

Over the past few decades, the $q$-state clock model has been extensively studied  \cite{joseRenormalizationVorticesSymmetrybreaking1977,elitzurPhaseStructureDiscrete1979,Cardy1980General, britoTwodimensionalClockModels2010,tobochnikPropertiesStateClock1982,challaCriticalBehaviorSixstate1986,yamagataPhaseTransitions6clock1991,tomitaProbabilitychangingClusterAlgorithm2002,borisenkoNumericalStudyPhase2011,britoTwodimensionalClockModels2010,krcmarPhaseTransitionSixstate2016,liAccurateSimulationQstate2020,chenPhaseTransitionState2017,liCriticalPropertiesTwodimensional2020,lapilli2006universality,hwang2009sixstate,baek2010comment,baek2010nonkosterlitzthouless,borisenko2012phase,baek2013residual,kumanoResponseTwistSystems2013,chatelainDMRGStudyBerezinskii2014,surunganBerezinskiiKosterlitzThouless2019,hongLogarithmicFinitesizeScaling2020,ortizDualitiesPhaseDiagram2012}, especially in two dimensions.
For $2 \leq q \leq 4$, the two-dimensional $q$-state clock model goes through a second-order phase transition from a low-temperature (low-$T$) ordered phase to a high-temperature (high-$T$) disordered phase as temperature increases. 
For larger $q$, the phase diagram is altered. In Ref.~\cite{elitzurPhaseStructureDiscrete1979} it was argued that there is a lower bound $q_{\rm  c}$ such that
for $q\ge q_{\rm c}$, a quasi-long-range ordered (QLRO) phase emerges, sandwiched by the ordered phase and the disordered phase, leading to two phase transition points, which are denoted by $\bco$ and $\bct$ ($\bco < \bct$).
To make the analysis tractable, the authors in Ref.~\cite{elitzurPhaseStructureDiscrete1979} performed a renormalization group (RG) analysis on the Villain clock model \cite{villain1975theory, joseRenormalizationVorticesSymmetrybreaking1977}, which still has $\mathbb{Z}_q$ symmetry and is assumed to be in the same universality as the clock model. 
They obtained that the lower bound is $q_{c}=5$ and the correlation function decays algebraically in the QLRO phase
\begin{equation}
    \label{eq:correlation_function}
	g(\mathbf{r}) = \langle {\bf{S}}_\mathbf{r} \cdot  {\bf{S}}_{\mathbf{0}}\rangle \sim \frac{1}{\|\mathbf{r}\|^{\eta(\beta)}},
\end{equation}
where the $\beta$-dependent exponent $\eta(\beta)$ varies from $\eta_1 = 1/4$ at $\bco$ to $\eta_2 = 4/q^2$ at $\bct$.
Moreover, they pointed out that both phase transitions are of Berezinskii--Kosterlitz--Thouless (BKT) type \cite{Berezinskii1971,kosterlitOrderingMetastabilityPhase,kosterlitzCriticalPropertiesTwodimensional1974}.

Even though the theoretical analysis predicts that both transitions belong to the BKT universality for $q\geq 5$, numerical confirmation seems 
to be rather tortuous. The authors of Ref.~\cite{lapilli2006universality} refuted the theoretical prediction and claimed the BKT transitions are numerically 
observed only for $q \geq 8$. Later, this statement was further supported by Ref.~\cite{hwang2009sixstate}, in which the Fisher zero approach was used to investigate the six-state clock model with the linear system size up to $L=28$ and found both phase transitions differ from the BKT transition.
However, Baek \emph{et al.} \cite{baek2010comment} studied the same model ($q=6$) with a much larger system size $L=512$ and claimed the phase transitions are of BKT type.  
Furthermore, Baek and Minnhagen    \cite{baek2010nonkosterlitzthouless} performed Monte Carlo simulations to study the helicity modulus of the five-state clock model and Villain clock model. 
For the five-state clock model, they observed two phase transitions, consistent with the theoretical result $q_{\rm c}=5$. However, they found that the helicity modulus remains finite for all temperatures and claimed the high-$T$ transition
is not of BKT type. On the other hand, the high-$T$ transition of the five-state Villain model was confirmed to be of the BKT type and the reason was ascribed to the residual $\mathbb{Z}_{5}$ symmetry of the clock model \cite{baek2013residual}. 
Later Kumano \textit{et al.} in Ref.~\cite{kumanoResponseTwistSystems2013} suggested that the definition of the helicity modulus for models with the discrete $\mathbb{Z}_{q}$ symmetry
should be modified. Based on the appropriately defined helicity modulus, they showed the existence of the two BKT transitions for $q\ge 5$.
This conclusion was further supported by various numerical methods \cite{borisenkoNumericalStudyPhase2011, borisenko2012phase, chatelainDMRGStudyBerezinskii2014,krcmarPhaseTransitionSixstate2016,liCriticalPropertiesTwodimensional2020, surunganBerezinskiiKosterlitzThouless2019,hongLogarithmicFinitesizeScaling2020}.

An important task is then to estimate the BKT phase transition points. G. Ortiz \textit{et al.} \cite{ortizDualitiesPhaseDiagram2012} invoked a bond-algebraic approach to demonstrate that the high-$T$ 
transition point $\bco \sim \rm{O}(1)$ and converges to the BKT transition point of the 2D $XY$ model as $q \to\infty$;
they also proved the low-$T$ transition point $\bct \sim q^2$. However, high-precision estimates of $\bco$ and $\bct$ are still challenging due to the multiplicative and additive logarithmic corrections in the finite-size-scaling behaviors near the critical points. 
Table~\ref{table: different works} lists some previous numerical results of the transition points for $q=5,6$ from the Monte Carlo (MC) and tensor network (TN) methods. 
Estimated results from different approaches are not entirely consistent and further investigation seems desired.

Another numerical interest of the clock model is to explore its duality property. For $2\le q \le 4$, the square-lattice model is known to be self-dual, 
and its unique critical point $\beta_{\rm c}$ is at its self-dual point $\beta_{\rm sd}$, which can be analytically derived (see Sec.~\ref{subsec:self-duality}).
For $q\geq 5$, the model is no longer strictly self-dual, but one can still obtain the self-dual point for $q=5$ by directly solving the
self-dual point equation, which corresponds to neither $\beta_{c1}$ nor $\beta_{c2}$. 
\begin{table}[t]
	\centering
	\caption{Estimated  critical points $\bco$ and $\bct$ for $q=5,6,7,8, 9$.
	Here, MC stands for the Monte Carlo method, and TN stands for the tensor network methods. 
	Our estimates are consistent with the previous MC results, with the precision being significantly improved. 
	Apart from some disagreements among themselves, several existing TN results are nearly excluded by our estimates 
	if the quoted error margins are taken seriously into account.}
	\begin{tabular}{llll}
	\hline\hline
$q$ & \multicolumn{1}{c}{Method and source}& \multicolumn{1}{c}{$\bco$} & \multicolumn{1}{c}{$\bct$} \\
	\hline
5	&Borisenko \textit{et al.} (2011, MC) \cite{borisenkoNumericalStudyPhase2011} & 1.0510(10)& 1.1048(10)\\
	&Kumano \textit{et al.} (2013, MC) \cite{kumanoResponseTwistSystems2013} & 1.059&1.101 \\
	&Chatelain (2014, TN) \cite{chatelainDMRGStudyBerezinskii2014}&1.06(2) &1.094(14)\\
	&Chen \textit{et al.} (2018, TN) \cite{chenPhaseTransitionsFivestate2018}&  1.050\,4(1)& 1.107\,5(1)  \\
	&Surungan \textit{et al.} (2019, MC) \cite{surunganBerezinskiiKosterlitzThouless2019}&1.064(6)&1.098(6) \\
	&Li \textit{et al.} (2020, TN) \cite{liCriticalPropertiesTwodimensional2020}&1.050\,3(2)&1.103\,9(2) \\
	&Hong and Kim (2020, TN) \cite{hongLogarithmicFinitesizeScaling2020}&1.058(1) &1.101(6)\\
	
	&Li \textit{et al.} (2020, TN) \cite{liAccurateSimulationQstate2020}&1.051\,9(6)&1.101(2) \\
	&\textbf{MC present work} & 1.055\,6(9)& 1.097\,5(6)  \\
	% \hline
6	&Tomita \& Okabe (2002, MC) \cite{tomitaProbabilitychangingClusterAlgorithm2002}&1.110\,1(7)&1.426(2)   \\
	&Brito \textit{et al.} (2010, MC) \cite{britoTwodimensionalClockModels2010}&1.11(1)&1.47(2) \\
	&Kumano \textit{et al.} (2013, MC) \cite{kumanoResponseTwistSystems2013}&1.106(6) &1.429(8) \\
	&Chen \textit{et al.} (2017, TN) \cite{chenPhaseTransitionState2017}&1.135\,8(3)&1.5020(11)\\
	&Surungan \textit{et al.} (2019, MC) \cite{surunganBerezinskiiKosterlitzThouless2019}&1.114(6)&1.43(1) \\
	&Li \textit{et al.} (2020, TN) \cite{liCriticalPropertiesTwodimensional2020}&1.095\,7(6)&1.449\,1(8) \\
	&Hong and Kim (2020, TN) \cite{hongLogarithmicFinitesizeScaling2020}&1.106(2)& 1.444(2)\\
	&Ueda \textit{et al.} (2020, TN) \cite{Ueda2020Finite}
	&1.101(4) & 1.441(6) \\
	&Li \textit{et al.} (2020, TN) \cite{liAccurateSimulationQstate2020}&1.097\,6(6)&1.437(4) \\
	&\textbf {MC present work}& 1.110\,3(15) & 1.427\,5(7)  \\
	% \hline
7	&Borisenko \textit{et al.} (2012, MC) \cite{borisenko2012phase} &1.111\,3(13) & 1.877\,5(75)\\
	&Chatterjee \textit{et al.} (2018, MC) 
	\cite{Chatterjee2018} & \multicolumn{1}{c}{-} & 1.88(2) \\
	&Li \textit{et al.} (2020, TN) \cite{liCriticalPropertiesTwodimensional2020}&1.102\,4(6)&1.8850(11) \\
	&Li \textit{et al.} (2020, TN) \cite{liAccurateSimulationQstate2020}&1.103\,1(6)&1.866(7) \\
	&\textbf {MC present work}& \multicolumn{1}{c}{-} & 1.851(1)  \\
	% \hline
8	&Tomita \& Okabe (2002, MC) \cite{tomitaProbabilitychangingClusterAlgorithm2002}&1.119\,1(9)&2.348(2)   \\
	&Li \textit{et al.} (2020, TN) \cite{liCriticalPropertiesTwodimensional2020}&1.103\,8(6)&2.396\,9(17) \\
	&Li \textit{et al.} (2020, TN) \cite{liAccurateSimulationQstate2020}&1.104\,9(6)&2.372(8) \\
	&\textbf {Our result (2022, MC)}& \multicolumn{1}{c}{-} & 2.349(2)  \\
	% \hline
9	&Li \textit{et al.} (2020, TN) \cite{liAccurateSimulationQstate2020}&1.104\,9(6)&2.924(17) \\
	&\textbf {MC present work}&1.119(2) & 2.920(2)\\
	\hline\hline
	\end{tabular}
	\label{table: different works}
\end{table}
On the other hand, there is no exact self-dual point for $q>5$. 
Nevertheless, some attempts have been made to obtain an approximate self-dual point.
In Ref.~\cite{ortizDualitiesPhaseDiagram2012} the authors proposed a variant of the clock model, which is exactly self-dual for 
all integer $q$ and related to the $q$-state clock model with $2\le q \le 4$. They argued that the self-dual point 
$\beta_{\rm sd}$ scales as $q/2\pi$ in the large $q$ limit. Later on, Chen \textit{et al.} \cite{chenPhaseTransitionState2017} used the TN method to define an approximate self-dual point via the 
normalized bond entanglement spectra in the original and dual lattices. Their $\beta_{\rm sd}$ values approximately scale as $q/2\pi + 1/4$ as $q \to \infty$.

The MC simulations of the $q$-state clock model to date work mainly on the standard spin representation.
% i.e., sampling spin configurations according to the corresponding probability distribution. 
There is, however, another way to investigate the system by formulating the model 
in terms of the closed-path (CP) configurations defined on bonds. A typical example that demonstrates the advantages of this transformation 
is the Ising model on a square lattice. In two dimensions, there are two ways of expressing the Ising model in terms of CP configurations, 
obtained via the high-$T$ expansion and low-$T$ expansion, respectively. 
The former expands the Boltzmann factor of each bond to decouple spins,
which can be generalized to higher dimensions; The latter keeps track of the domain-wall boundaries on the dual lattice. The CP configurations in the two expansions can be sampled by the worm algorithm \cite{prokofevWormAlgorithmQuantum1998,prokofevWormAlgorithmsClassical2001,dengDynamicCriticalBehavior2007, hitchcockDualGeometricWorm2004,elciLiftedWormAlgorithm2018}, which is at least as efficient as 
cluster algorithms for the spin representation. Moreover, it is very convenient to measure the two-point correlation function in the worm simulation. 
Finally, the representations obtained from the high-$T$ and low-$T$ expansions provide a natural way to study the duality of the model. 
For the 2D Ising model, the two expansions are used to derive the Kramers-Wannier duality \cite{kramersStatisticsTwoDimensionalFerromagnet1941}.
Both expansions can be applied to a broad class of lattice models \cite{parisi1988statistical}. In Ref.~\cite{wangPercolationTwodimensionalModel2021}
the high-$T$ CP formulation was applied to the 2D $XY$ model, where the bond variables now take
integer values and obey the Kirchhoff conservation laws. Because of the resemblance of the bond variables to flows, 
this representation is also called the flow representation.

\begin{table}[H]
	\centering
	\caption{Summary of the final estimates of the critical point $\bco$ and $\bct$ and the self-dual point $\beta_{\rm sd}$ with  $q=5,6,7,8,9$. 
	Also shown are the results for the anomalous dimension $\eta(\beta_{\rm sd})$ at the self-dual point, which are equal to $1/q$ within the error margin. 
	For $q \geq 6$, the values of $\beta_{\rm c1}$ are very close and differ only at the third decimal place, having the same magnitude as the error margins, 
	and thus, we do not determine $\beta_{\rm c1}$ for $q=7,8$. }
	\begin{tabular}{clllll}
		\toprule
 		$q$ & \multicolumn{1}{c}{$\bco$} & \multicolumn{1}{c}{$\bct$} & \multicolumn{1}{c}{$\beta_{\rm sd}$} &\multicolumn{1}{c}{$\eta(\beta_{\rm sd})$} &\multicolumn{1}{c}{$1/q$} \\
		\cmidrule[.4pt](r){1-4} \cmidrule[.4pt](l){5-6}
		5 &~~1.0556(9)& 1.097\,5(6)  & 1.076\,318\dots &~~0.200(2)  &~~0.2\\
		6 &~~1.110\,3(15)& 1.4275(8)  & 1.254\,10(5) &~~0.166\,5(3) &~~0.1667$\dots$\\
		7 & \multicolumn{1}{c}{-} & 1.851(1)  & 1.417\,11(6) &~~0.142\,6(7) &~~0.1429$\dots$\\
		8 & \multicolumn{1}{c}{-} & 2.349(2)  & 1.573\,06(9) &~~0.125\,0(2) &~~0.125\\
		9 &~~1.119(2) & 2.920(2) & 1.727\,18(9) &~~0.111\,1(2)              &~~0.1111$\dots$\\
		\bottomrule
        % \hline 
	\end{tabular}

	\label{tab: summary}
\end{table}

\begin{figure}[t]
	\centering
	\includegraphics[width=.8\columnwidth]{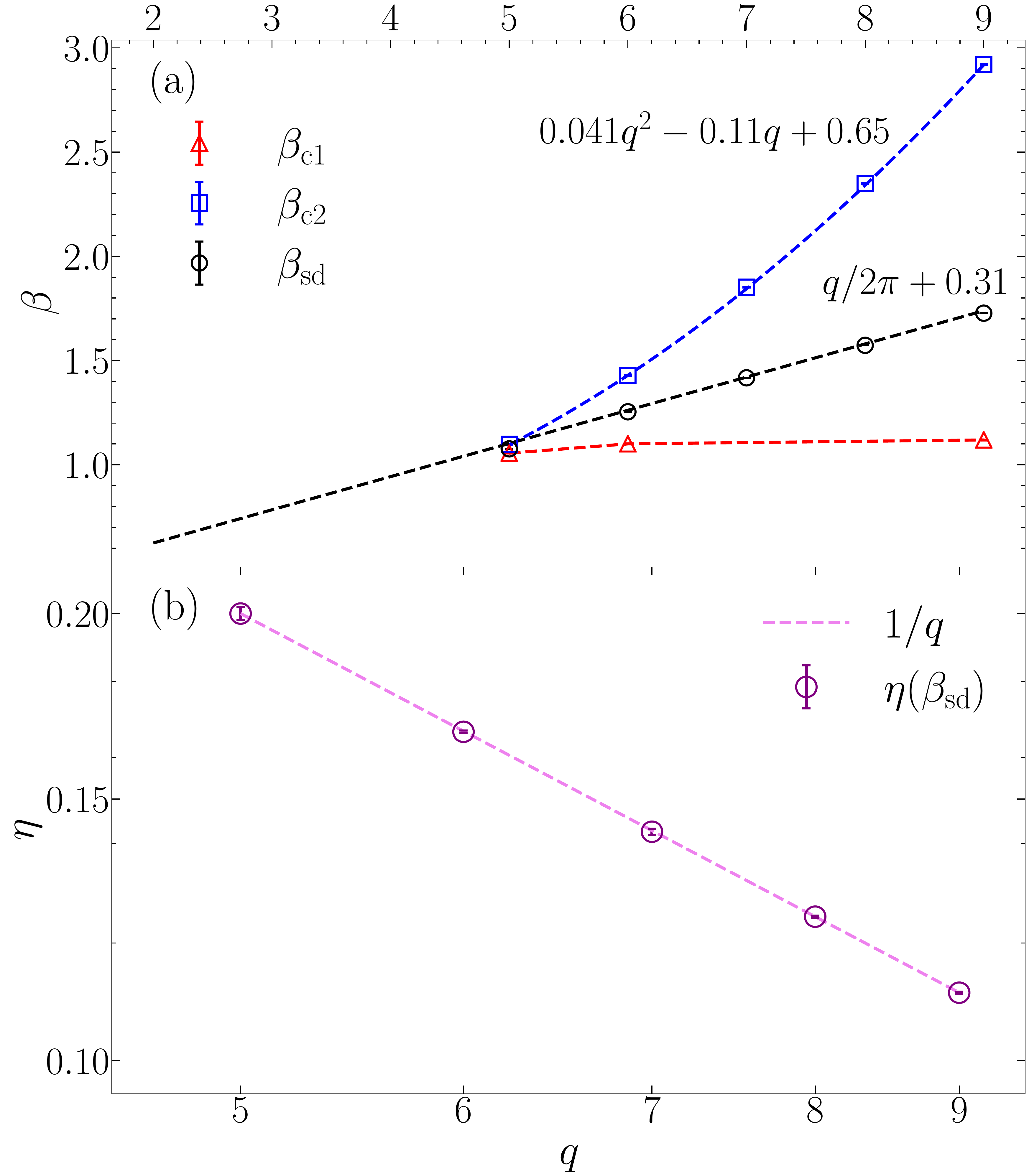}
	\caption{(a) Plot of $\bco, \bct,$ and $\beta_{\rm sd}$ versus $q$. The blue dashed line represents $\bct(q) = 0.041q^2 - 0.11q + 0.65$ 
	and the black dashed line represents $\beta_{\rm sd}(q) = q/2\pi + 0.31$. (b) Anomalous dimension $\eta(\beta_{\rm sd})$ at the self-dual point 
	as a function of $q$ on a log-log scale. The data points can be fitted accurately by the function $1/q$.
	}
	\label{fig: SDKc}
\end{figure}
In this work we study the flow representations of the 2D $q$-state clock model derived from the high-$T$ and low-$T$ expansions.
Different from the flow representations of the two-dimensional $XY$ model, the bond variables take values from $\{0,\dots,q-1\}$ 
and satisfy the modified flow conservation law, which we call the $q$-modular flow conservation (see Sec.~\ref{Sec: Expansion}). 
We formulate a worm algorithm to efficiently simulate the clock model in both representations and perform extensive simulations with $5\leq q \leq 9$ and linear system size up to $L=1024$. In our simulations, we measure the average value of the worm returning time in the high-$T$ and low-$T$ flows,
denoted by $\chih$ and $\chil$, respectively. It can be proved that $\chih$ is strictly equal to the magnetic susceptibility in the spin representation, whose critical behavior is already known.
As for $\chil$, based on the duality between the two flow representations, we expect it to exhibit a dual scaling behavior to $\chih$. % corresponds to the susceptibility $\chi_\textsc{h}$ for the high-$T$ flows and $\chi_\textsc{l}$ for the low-$T$ flows.
By performing finite-size analysis of the data of both quantities,  we get the estimates of $\bco$ from $\chi_\textsc{h}$ and $\bct$ from $\chi_\textsc{l}$.
The estimates of $\bco$ and $\bct$ are summarized in Table~\ref{tab: summary}.
In Fig.~\ref{fig: SDKc}(a) we plot $\bco$ (red triangles) and $\bct$ (blue squares) versus $q$.
 As Table~\ref{tab: summary} and Fig.~\ref{fig: SDKc} demonstrate, the low-$T$ transition point $\bct$ scales as $\bct(q) = a_0 + a_1q + a_2q^2$ with $a_0 = 0.65(6), a_1 = -0.11(2)$, and $a_2=0.041(1)$. A consistent leading scaling behavior was observed in Ref.~\cite{borisenko2012phase}. For the high-$T$ transition point $\bco$, it quickly approaches the 2D $XY$ model transition point $\beta_{\rm BKT} = 1.119\,96(6)$ \cite{komuraLargescaleMonteCarlo2012} as $q$ increases.

The precision of our estimates is significantly greater than previous MC results.
Furthermore, our estimates nearly exclude a number of the existing TN results. To be more specific, for two estimates $\beta_1$ and $\beta_2$ with error margins $\sigma_1$ and $\sigma_2$, we consider them inconsistent if $|\beta_1 - \beta_2| > 3\sigma_1 + 3\sigma_2$. According to this criterion, estimates of $\bco$ and $\bct$ in Refs.~\cite{chenPhaseTransitionsFivestate2018,liCriticalPropertiesTwodimensional2020} for $q=5$,
$\bco$ in Refs.~\cite{chenPhaseTransitionState2017,liCriticalPropertiesTwodimensional2020,liAccurateSimulationQstate2020} and $\bct$ in Refs.~\cite{chenPhaseTransitionState2017,liCriticalPropertiesTwodimensional2020,hongLogarithmicFinitesizeScaling2020} for $q=6$, $\bct$ in Ref.~\cite{liCriticalPropertiesTwodimensional2020} for $q=7, 8$, and $\bco$ in Ref.~\cite{liAccurateSimulationQstate2020} for $q=9$ are unlikely.
The deviations of these results are probably because, due to the cutoff of bond dimension in the TN calculation, systematic biases are unavoidably introduced 
but difficult to estimate reliably, particularly near the critical points where the correlation length is divergent.

\begin{figure}[t]
	\centering
	\includegraphics[width = \columnwidth]{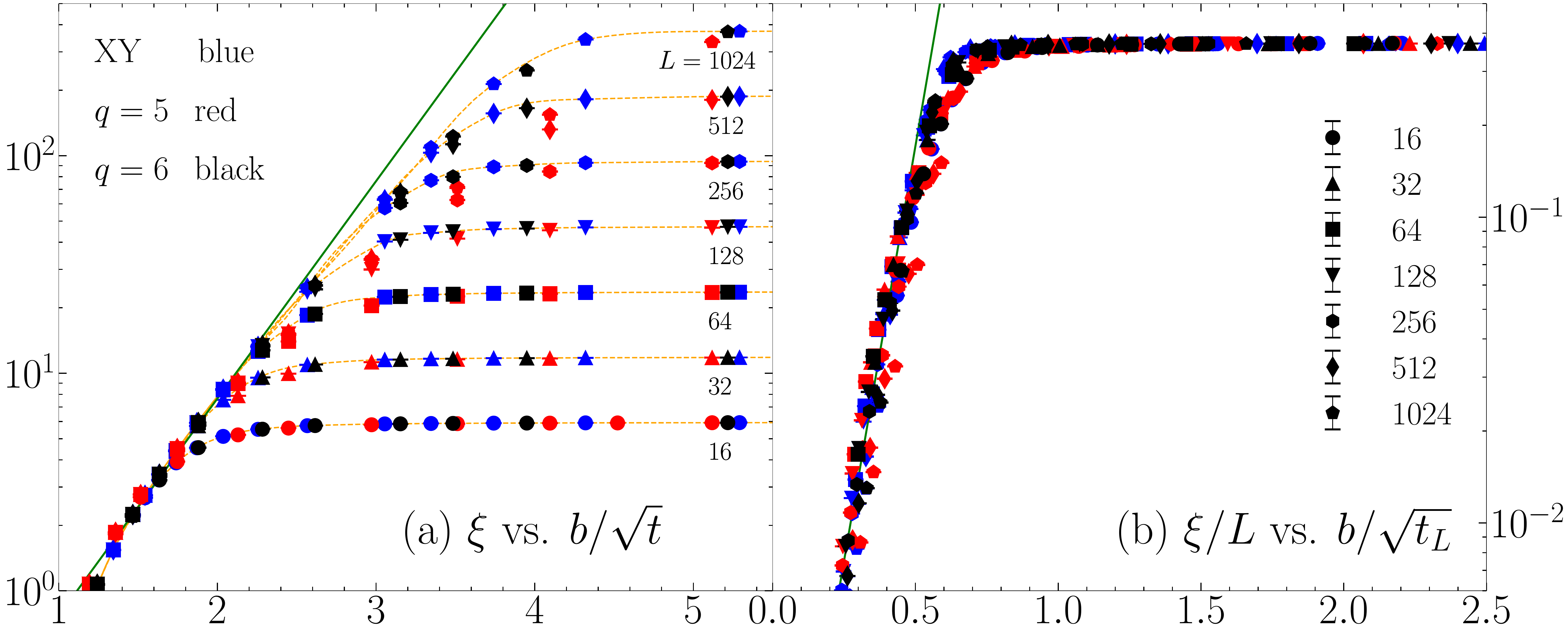}
	\caption{(a) Semilogarithmic plot of the correlation length $\xi$ versus $b/\sqrt{t}$ for the 2D XY, $q=5, 6$ clock model, where $t = (\bco - \beta)/\bco$, illustrating the exponential growth of $\xi$.
	Different models are distinguished by different colors, as defined in the legend. The corresponding nonuniversal constant $b = 1$ (XY), 0.94 ($q=5$), 0.994 ($q=6$).
	(b) Semilogarithmic plot of the ratio $\xi / L$ as a function of $b/\sqrt{t_L}$, where $t_L = t[\ln (L/L_0)]^2$ with the nonuniversal length scale simply set by $L_0 =1 $. The data points collapse onto a single curve, indicating $\xi/L$
	is a universal function of the scaling field $b/\sqrt{t_L}$. These plots clearly support that the $q=5$ and 6 clock models are in the universality class of the $XY$ model. 
	In particular, the resemblance of the three systems is clearly illustrated by the approximately identical values of $b$; actually, the $b$ value for $q=6$ is hardly different from that for the 2D $XY$ model, which is a rather surprising observation.}
	\label{fig:xi_beta1}
\end{figure}

To demonstrate that $\bco$ is of BKT type for $q=5, 6$, we measure the correlation length $\xi$, which is defined as
\begin{equation}
	\xi = \frac{\int  \|\mathbf{r}\| g(\mathbf{r}) d\mathbf{r}}{\int g(\mathbf{r}) d\mathbf{r}},
\end{equation}
where $g(\mathbf{r})$ is the two-point correlation function. In the worm algorithm, this quantity can be evaluated by averaging the distance between the two defects $\scrI$ and $\scrM$, 
which are introduced to study the two-point correlation function (see Sec.~\ref{Sec: Algorithm}). For a BKT transition, $\xi$ diverges according to the asymptotic law $\xi \sim \exp(b/\sqrt{t})$ \cite{kosterlitOrderingMetastabilityPhase,kosterlitzCriticalPropertiesTwodimensional1974}, with $t = (\bco - \beta)/\bco$ and $b$ a nonuniversal constant. 
In Fig.~\ref{fig:xi_beta1}(a), 
we plot the correlation length $\xi$ versus $b/\sqrt{t}$ on a semilogarithmic scale for the clock model with $q=5,6$ and the 2D $XY$ model.
The approximately linear behavior of the curves when $\xi \ll L$ implies that the correlation length diverges exponentially with $b/\sqrt{t}$.
This exponential growth, together with its resemblance to the curve of the $XY$ model, directly confirms that $\bco$ is a BKT transition for $q=5, 6$. 
Further, this exponential scaling behavior, which should in principle be valid at large scales only, appears even when $\xi$ is comparable to the microscopic scale (lattice spacing). When $\xi$ is 
comparable to the system size, the curves enter a region of plateaus due to the finite-size cutoff.
We further plot the ratio $\xi/L$ versus $b/\sqrt{t_L} = b/\sqrt{t(\ln L/L_0)^2}$ in Fig.~\ref{fig:xi_beta1}(b), where $L_0$ is some nonuniversal characteristic length of the order of the lattice constant. The scaling field $t_L = t (\ln L/L_0)^2$ \cite{Harada1997Universal,kumanoResponseTwistSystems2013} originates from the exponential divergence of the correlation length near the BKT transition $\xi \sim L \sim \exp(b/\sqrt{t})$.
We adjust the nonuniversal constants $b$ and $L_0$ such that the data of different models and system sizes collapse, and find that simply setting $L_0=1$ is sufficient to demonstrate the approximate data collapse. 
There are some noticeable finite-size corrections for $q=5$, which is probably because $\bco$ and $\bct$ are too close for $q=5$ such that the scaling behavior of $\xi$ is also affected by $\bct$.
The corresponding values of $b$ are 0.94 ($q=5$), 0.994 ($q=6$), and 1 (XY). Surprisingly, these nonuniversal constants are very close to each other, and the $b$ value of the six-state clock model is nearly indistinguishable from that of the 2D $XY$ model. These strongly support the same universality of the five- and six-state clock models and the $XY$ model.

In the spin representation, the spin-spin correlation function is governed by the anomalous dimension $\eta_1=1/4$ at $\bco$ and $\eta_2=4/q^2$ at $\bct$.
This is confirmed by the finite-size scaling of $\chih$ at the two critical points.
By studying the high-$T$ and low-$T$ flows, we further observe the simultaneous existence of
the pair of exponents $(\eta_1, \eta_2)$ at each BKT transition point.
At $\bco$, we have $\chi_\textsc{h}(\bco, L) \sim L^{2-\eta_1}$ and $\chi_\textsc{l}(\bco, L) \sim L^{2-\eta_2}$,
and vice versa at $\bct$. 
Moreover, we find that $\chil(\bco,L)$ is nearly identical to $\chih(\bct,L)$ for each system size $L$ and any given $q$. 
Similarly, we have $\chih(\bco,L) \approx \chil(\bct,L)$. These numerical results vividly demonstrate the duality between the two critical 
points and between the two flow representations. 

Taking advantage of the two dual flow representations, we define an approximate self-dual point for $q>5$ via the stringent condition 
$\chi_\textsc{h}(\beta_{\rm sd}, L)=\chi_\textsc{l}(\beta_{\rm sd}, L)$, 
% in which we not only require $\chih$ and $\chil$ to have the same scaling form but also the same amplitude.
in which we require both the scaling behaviors and amplitudes of $\chih$ and $\chil$ to be the same.
The self-dual point $\beta_{\rm sd}(L)$ for finite systems is found to be nearly independent of system size $L$.
To estimate $\beta_{\rm sd}$ in the thermodynamic limit, we perform the least-squares fit to the data of 
$\chi_{\rm diff} (\beta,L) = \chi_\textsc{h} (\beta,L) - \chi_\textsc{l}(\beta,L)$ and get estimates of $\beta_{\rm sd} \equiv \lim_{L\to \infty} \beta_{\rm sd}(L)$ for $q=6$--9. 
The results are summarized in Table~\ref{tab: summary}, which can be fit by the expression 
$\beta_{\rm sd}(q) = q/2\pi + 0.31(1)$, as shown in Fig.~\ref{fig: SDKc}. At our estimated $\beta_{\rm sd}$, we extract the exponent $\eta(\beta_{\rm sd})$
and find that $\eta(\beta_{\rm sd})$ is in excellent agreement with $1/q$, as shown in Fig.~\ref{fig: SDKc}(b), which can be understood from the perspective of the Villain clock model (see Sec.~\ref{sec:Villain}). Based on this, we conjecture that $\eta(\beta_{\rm sd})=1/q$ holds exactly at our defined self-dual point.

The remainder of this paper is organized as follows. Section~\ref{Sec: Expansion} introduces the flow representation,
studies the duality properties of the clock model, and summarizes the RG analysis of the Villain clock model. Section~\ref{Sec: Algorithm} describes the worm algorithm and sampled quantities.
In Sec.~\ref{Sec: Results}, the MC data are analyzed, and the results are presented. A brief summary is given in Sec.~\ref{Sec:Conclusion}.

\section{Flow representations, Duality, and RG analysis}
\label{Sec: Expansion}
In this section we elaborate on the derivations of the high-$T$ and low-$T$ flow representations for the convenience of general readers.
This technique can be applied to a broad class of models, such as the Potts model. On the basis of these two expansions, we study the duality and self-duality properties of the $q$-state clock model. Finally, we summarize the RG analysis of the Villain clock model, 
which should be in the same universality class as the original clock model.

\subsection{High-temperature expansion}
\label{subsec:HT_expansion}
\begin{figure}[h]
	\centering
	\includegraphics[width=\columnwidth]{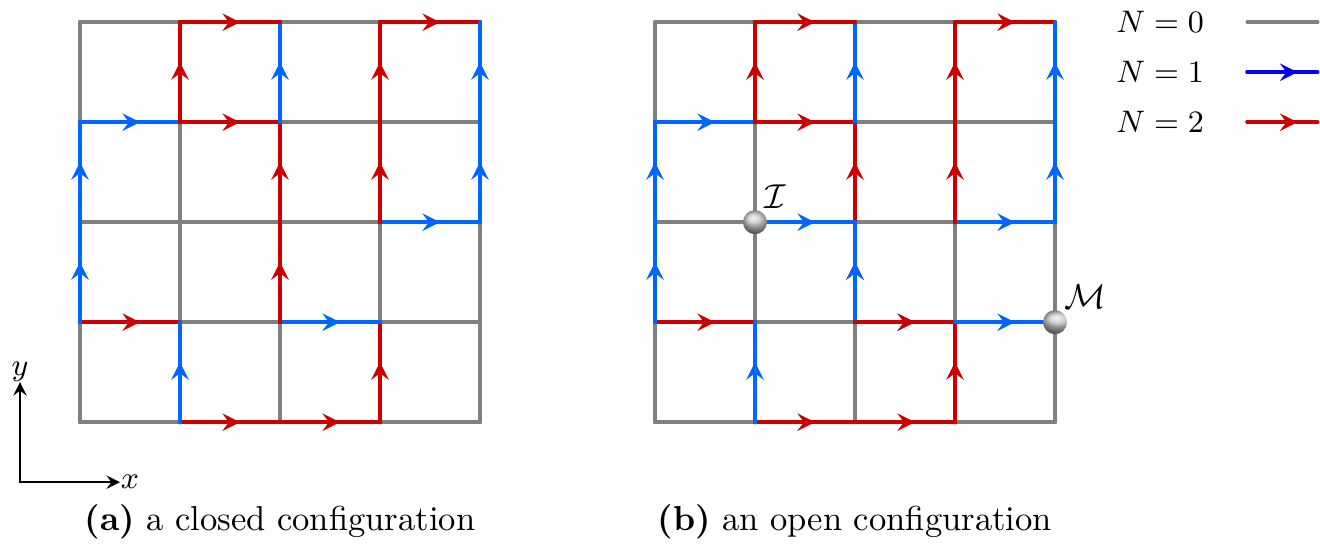}
	\caption{Illustration of two types of flow configurations in the high-$T$ expansion for $q=3$:
	(a) closed configuration with $(\nabla \cdot N)\bmod q = 0$ and (b) open configuration with defects $\scrI$ and $\scrM$ (gray circles) violating the q-modular flow conservation.
	The direction of horizontal (vertical) edges of $G$ is specified by the positive direction of the $x$axis ($y$axis).
	}
	\label{fig:HT-flows}
\end{figure}

Let $G \!=\! (V,E)$ denote a graph with $|V|$ vertices and $|E|$ edges. For each edge $\langle ij \rangle \in E$,
because the Boltzmann factor $f(\sigma_i - \sigma_j) = \exp{\beta \cos[\frac{2\pi}{q} (\sigma_i - \sigma_j)]}$ is a periodic function of variable $(\sigma_i - \sigma_j)$ with period $q$,
we can expand it into discrete Fourier series as
\begin{align}
	&f(\sigma_i - \sigma_j) = \sum_{N_{ij} = 0}^{q-1} F(N_{ij}) e^{2\pi \mathrm{i} (\sigma_i - \sigma_j)N_{ij}/q }, \label{eq:DFT}\\
    &F(N_{ij}) = \frac{1}{q}\sum_{\sigma_i - \sigma_j = 0}^{q-1}  f(\sigma_i - \sigma_j) e^{-2\pi \mathrm{i} (\sigma_i - \sigma_j) N_{ij}/q}, \label{eq:inv_DFT} 
\end{align}
where $N_{ij} \in \{0, \dots, q-1\}$ is the bond variable defined on the edge $\langle ij \rangle$, and $F(N_{ij})$ is derived from the inverse discrete Fourier transform~\eqref{eq:inv_DFT}. 
Plugging the expansion~\eqref{eq:DFT} into Eq.~\eqref{eq:partition_function}, we rewrite the partition function $\scrZ$ as
\begin{align}\label{eq:HT_expansion}
	\scrZ &= \sum_{\{\sigma\}}  \prod_{\langle ij \rangle } f(\sigma_i -\sigma_j) \nonumber\\
    &= \sum_{\{\sigma\}} \prod_{\langle ij \rangle} \left[\sum_{N_{ij} = 0}^{q-1} e^{2\pi \mathrm{i} (\sigma_i-\sigma_j)N_{ij}/q}F(N_{ij})\right] \nonumber\\
    &= \sum_{\{N\}} \sum_{\{\sigma\}}  \left[\prod_{\langle ij \rangle} F(N_{ij}) \! \cdot \! \prod_{\langle ij \rangle} e^{2\pi \mathrm{i}  (\sigma_i-\sigma_j)N_{ij}/q}\right] \nonumber\\
    &= \sum_{\{N\}}  \left[\prod_{\langle ij \rangle} F(N_{ij}) \! \cdot \!  \prod_{i} \!  \sum_{\sigma_i=0}^{q-1} \!  e^{2 \pi \mathrm{i} (\nabla \cdot \mathbf{N})_i \sigma_i/q}\right]  \nonumber\\
    &= q^{|V|} \sum_{\{N\}: \nabla \cdot \mathbf{N} = 0} \prod_{\langle ij \rangle} F(N_{ij}). 
 \end{align}
 % \end{equation}
% In the third equality, we use the fact that $\prod_{\langle ij \rangle} \sum_{N_{ij}} = \sum_{\{N\}} \prod_{\langle ij \rangle}$.
Here, $\sum_{\{N\}}$ sums over all configurations of bond variables $\{N\}$
and $(\nabla \cdot \mathbf{N})_i = \sum_{j:\langle ij \rangle \in E} \text{sgn}(i \! \to \! j) N_{ij}$ represents the divergence of $\{N\}$ at site $i$.
We specify a positive direction for each edge, and $\text{sgn}(i\to j)$ is +1 if the direction $i\to j$ aligns with the positive direction
and -1 otherwise. The sign function is introduced because $N_{ij}$ gives opposite contributions to $(\nabla \cdot \mathbf{N})_i$ and $(\nabla \cdot \mathbf{N})_j$.
In the last equality, we use the following identity to integrate out the spin variables $\{\sigma\}$
\begin{equation}
	% \sum_{\sigma_i = 0}^{q-1} \exp(\frac{2\pi \mathrm{i}}{q}(\nabla \cdot \mathbf{N})_i \sigma_i) = 
	\sum_{\sigma_i = 0}^{q-1} e^{2\pi \mathrm{i}(\nabla \cdot \mathbf{N})_i \sigma_i/q} = 
	\begin{cases}
	q & \text{for } (\nabla \cdot \mathbf{N})_i \bmod q = 0\\
	0 & \text{otherwise.}
	\end{cases}
\end{equation}
Therefore, only flows satisfying the $q$-modular flow conservation, i.e., $(\nabla \cdot \mathbf{N})_i \bmod q = 0$ for all $i \in V$, have nonzero
statistical weights. Graphically, this condition requires flows to form closed loops, and we use $\nabla \cdot \mathbf{N} = 0$ as shorthand for the condition. 
Equation~\eqref{eq:HT_expansion} is called the high-$T$ expansion of the clock model. Note that the final expression is independent of the choice of positive direction because every choice ensures that $N_{ij}$ contributes oppositely to its two endpoints.
With the above definition, $\text{sgn}(i \to j)N_{ij}$ can be regarded as a directed flow defined on $E$, hence the name of flow representation.
Figure~\ref{fig:HT-flows}(a) shows a closed flow configuration of the $q=3$ clock model in the high-$T$ expansion. The value of the bond variable $N_{ij}$ 
is specified by the color of the edge and the positive directions of the flows are specified by the arrows. 

\subsection{Low-temperature expansion}
\label{subsec:LT_expansion}
In addition to the high-$T$ expansion, which is applicable to any spatial dimensions, there is another flow representation of the model in two dimensions via the low-$T$ expansion, 
which utilizes the dual lattice $G^* = (V^*, E^*)$. For a given planar lattice $G$, its dual lattice $G^*$ can be formed as follows: 
(i) On the center of each face of $G$, place a vertex that serves as the dual vertex; (ii) for any two vertices of $G^*$, add an edge between them if the corresponding two faces of $G$ have a common edge. 
As a result, there is a one-to-one correspondence between the edges of $G$ and $G^*$.

\begin{figure}[h]
	\centering
	\includegraphics[width=.8\columnwidth]{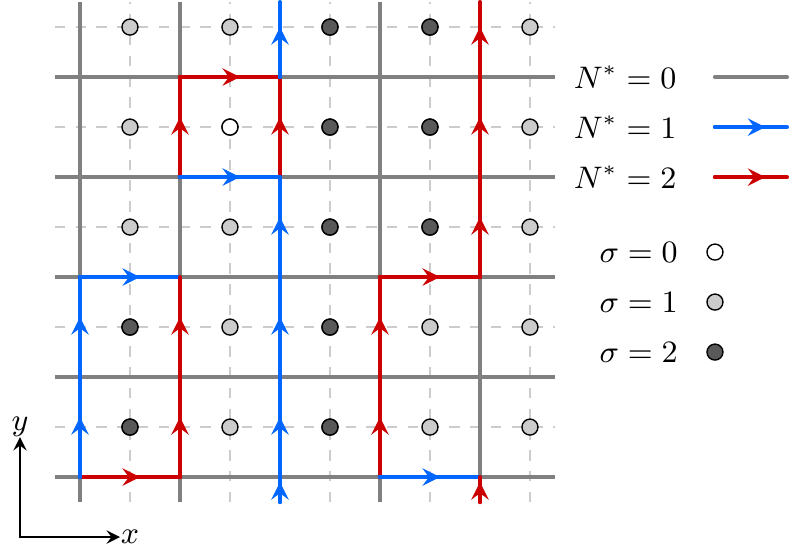}
	\caption{Illustration of a closed flow configuration in the low-$T$ expansion for $q=3$ with periodic boundary conditions. 
	The closed circles and dashed gray lines constitute the original lattice $G$. The dual vertices are at the faces of $G$,
	with dual edges connecting them. The direction of the horizontal (vertical) edges of $G^*$ is specified by the positive direction of the $x$axis ($y$axis). The value of the flow variable on each edge is determined by the difference between its right and left spin.
	}
	\label{fig:LT_flows}
\end{figure}

Similar to the high-$T$ expansion, we specify a positive direction for each edge in $E^*$ and introduce a new set of bond variables defined on $E^*$ as $N_{ij}^* = (\sigma_r - \sigma_l) \bmod{q}$, with $\sigma_r$ and $\sigma_l$ denoting the clock spins on the right and left sides of the 
positive direction of $\langle ij\rangle$ in $E^*$. By definition, $N_{ij}^*$ also takes integer values in the range $[0,q-1]$. 
Analogously, the divergence of $\{N^*\}$ at site $i\in V^*$ is defined as $(\nabla \cdot \mathbf{N}^*)_i = \sum_{j:\langle ij \rangle \in E^*} \text{sgn}(i\to j) N_{ij}^*$, 
which automatically satisfies the $q$-modular flow conservation $(\nabla \cdot \mathbf{N}^*)$ mod $q =0$. The partition function therefore can be rewritten as

\begin{align}
	\scrZ &= \sum_{\{\sigma\}} \prod_{\langle ij \rangle \in E} \exp\left [\beta \cos\left(\frac{2\pi}{q}(\sigma_i-\sigma_j)\right)\right] \nonumber \\
	&= q\sum_{\{N^*\}: \nabla \cdot \mathbf{N}^* = 0} \left(\prod_{\langle ij \rangle\in E^*} F^*(N^*_{ij})\right), \label{eq:LT_expansion}
\end{align}
where $ F^*(N^*_{ij}) = \exp\left[\beta\cos(\frac{2\pi}{q}N^*_{ij})\right]$ is a periodic function with period $q$. 
The factor $q$ originates from the $q$-to-one correspondence between spin configurations and flow configuration (global $\mathbb{Z}_q$ symmetry).
% Figure~\ref{fig:LT_flows} gives an example of a flow configuration of the$q=3$ clock model on the square lattice, which constructs the low-$T$ flows from a spin configuration and consists of closed loops in the dual lattice.
Figure~\ref{fig:LT_flows} illustrates a closed flow configuration of the $q=3$ clock model in the low-$T$ expansion, where the low-$T$ flows are constructed from a spin configuration and form closed loops in the dual lattice.
The edges of the original lattice are represented by dashed lines, and spins with different values are distinguished by their colors. 
The solid lines consist of the edges of the dual lattice, where the values of the flows are specified by the colors of the edges.

\subsection{Duality of the 2D $q$-state clock model}
\label{subsec:self-duality}
The flow representations derived from the high-$T$ and low-$T$ expansions provide a convenient way to study the duality property of the clock model.
To begin with, we first define the ratios $R(\beta,N) \equiv F(\beta,N)/F(\beta,0)$ for the high-$T$ expansion, which characterizes the relative weight of bond with value $N$. 
For any given $N > 0$, the ratio $R(\beta, N)$ is a monotonically increasing function of $\beta$ and satisfies $0 < R(\beta, N) < 1$.
% For all $N > 0$, the ratios $0 < R(\beta, N) < 1$, and $R(\beta,N)$ is a monotonically increasing function of the inverse temperature $\beta$ for each fixed $N$. 
In the high-$T$ limit, i.e., $\beta \to 0$, we have $R(\beta,N) \ll 1$, which means that $N=0$ has a much larger statistical 
weight than that of other possible values of $N$, and configurations with dilute loops dominate in this case. As temperature decreases, 
the weights of nonzero bond values begin to increase and the loop gases become denser.
Likewise, for the low-$T$ expansion, we define the ratio $R^*(\beta,N) \equiv F^*(\beta,N)/F^*(\beta,0)$, which also takes value ranging from 0 to 1 but
is a monotonically decreasing function of $\beta$. In the low-$T$ limit, i.e., $\beta \to \infty$, 
we have $R^*(\beta,N) \ll 1$; therefore, most of the bonds in the low-$T$ expansion now have value 0, forming dilute loop gases. 
As temperature increases, the ratio $R^*(\beta, N)$ becomes larger and the loop-density increases.

From the above analysis, we can see that, as temperature decreases, the low-$T$ flows undergo an opposite process as the high-$T$ flows.
Thus, one may expect that there exists a correspondence between the high-$T$ expansion at $\beta$ and the low-$T$ expansion at $\beta^*$ from the perspective of the
loop distribution, which implies the following dual equation set
\begin{equation} \label{eq:self_dual}
    R(\beta,N) =  R^*(\beta^*,N)
    \quad (N=1,\dots,q-1),
\end{equation}
where $R$ is defined for the high-$T$ flows on the original lattice $G$ while $R^*$ is defined for the low-$T$ flows on the dual lattice $G^*$.
The $q-1$ equations in \eqref{eq:self_dual} are not independent due to the relations
$R(\beta, N) = R(\beta, q-N)$ and $R^*(\beta^*, N) = R(\beta^*, q-N)$, hence there are at most $\lfloor q/2 \rfloor$ independent equations, with $\lfloor q/2 \rfloor$ the integer part of $q/2$. 
Furthermore, for self-dual lattices, i.e., the original lattice and the dual lattice have the same geometry, if there exists
a function $\beta^*(\beta)$ such that the set of equations~\eqref{eq:self_dual} are satisfied, we say the model is self-dual in the sense that the theory at $\beta$
is related to itself at $\beta^*$.

For self-dual models, one can further set $\beta^* = \beta = \beta_{\rm sd}$, which gives the equation set for the self-dual temperature $\beta_{\rm sd}$
\begin{equation} \label{eq:self_dual_point}
    R(\beta_{\rm sd},N) =  R^*(\beta_{\rm sd},N)  \quad (N=1,\dots,q-1).
\end{equation}
If the self-dual model only has one phase transition point $\beta_c$, then $\beta_{\rm sd} = \beta_c$. Otherwise, one can find more than one point at which the free energy exhibits singularity. 
For the case of two phase transition points $\bco$ and $\bct$, one may expect that $R(\bco, N) = R^*(\bct, N)$ and $R(\bct, N) = R^*(\bco, N)$.

Now let us consider the cases of $q=2,3,4$ for a 2D square lattice, which correspond to the Ising model, three-state Potts model, and two copies of the Ising model, respectively.
The corresponding duality conditions~\eqref{eq:self_dual} are
\begin{align}
	q=2: & \quad \tanh(\beta) = e^{-2\beta^*}, \quad  \quad \ \,N=1;\\
	q=3: & \quad \frac{e^{3/2\beta}-1}{e^{3/2\beta}+2} = e^{-3/2\beta^*}, \quad N=1,2;\\
	q=4: & ~\begin{cases}
	\tanh(\beta/2) = e^{-\beta^*}, & N=1,3, \\
	\tanh^2(\beta/2) = e^{-2\beta^*}, &N=2.
	\end{cases}
\end{align}
% \end{equation}
In each case, given a $\beta$, there is a unique $\beta^*$ satisfying the equation set. Therefore, the 2D $q$-state clock on a square lattice
model is self-dual for $q=2,3,4$. In particular, the self-dual point is calculated by setting $\beta^* = \beta = \beta_{\rm sd}$, which gives
\begin{equation}
	\beta_{\rm sd} = \begin{cases}
		\frac{1}{2}\ln (\sqrt{2} + 1), & q=2 \\
		\frac{2}{3}\ln(\sqrt{3} + 1),  & q=3 \\
		\ln(\sqrt{2}+1),			   & q=4.	
	\end{cases}
\end{equation}
Since there is only one phase transition point, the self-dual point is also its critical point. Note that $\beta_{\rm sd}(q=4) = 2\beta_{\rm sd}(q=2)$, which comes from the fact that the $q=4$ clock model is equivalent to two decoupled Ising models ($q=2$), i.e., $\scrZ_{q=4}(\beta) = \scrZ_{q=2}^2(\beta/2)$.

For $q=5$, the dual equations are overdetermined in general, meaning there is no consistent solution of $\beta^*$ as a function of $\beta$ for arbitrary $\beta$.
Nevertheless, if we try to get the self-dual point by solving the set of equations~\eqref{eq:self_dual_point} for $\beta_{\rm sd}$, we get 
a single independent equation for $q=5$:
\begin{equation}
\frac{e^{5 \beta_{\rm sd} / 4}}{\cosh \left(\sqrt{5} \beta_{\rm sd} / 4\right)}=\sqrt{5}+1,
\end{equation}
which has the numerical solution $1.076\,318\,0\cdots$. At this particular point, the thermodynamic properties of the low-$T$ flows
and high-$T$ flows are the same, but $\beta_{\rm sd}$ does not correspond to any of the critical points \cite{chenPhaseTransitionState2017} since the system now has two critical temperatures. 
For $q\geq 6$, there is neither a solution for the dual temperature $\beta^*(\beta)$ nor a solution for the self-dual point $\beta_{\rm sd}$. 
Nevertheless, given the similar physical pictures of high-$T$ and low-$T$ flows, we expect that there is still a duality between the low-$T$ and high-$T$ flows from the viewpoint of universality class, 
as we explore in Sec.~\ref{Sec: Results}.

\subsection{RG analysis of the Villain clock model}
\label{sec:Villain}
In this section, we briefly summarize the analytical results derived from the Villain clock model for $q\geq 5$, which sheds light on the original model since it has been demonstrated \cite{joseRenormalizationVorticesSymmetrybreaking1977,kadanoffLatticeCoulombGas1978,nienhuisCriticalBehaviorTwodimensional1984,tomitaProbabilitychangingClusterAlgorithm2002,surunganBerezinskiiKosterlitzThouless2019} 
that the Villain clock model and original clock model belong to the same universality class.
% that the Villain clock model has the same phase structure and critical exponents as the $q$-state clock model. 
The Villain model is obtained by replacing the Boltzmann factor $f(\sigma_i - \sigma_j)$ of the original clock model with a periodic Gaussian function
\begin{equation}\label{eq:Gaussian}
    % f_{\textsc{v}}(\sigma_i - \sigma_j)=\sum_{m=-\infty}^{\infty} \exp \left[-\frac{1}{2} \beta(\frac{2\pi}{q} (\sigma_i - \sigma_j)-2 \pi m)^{2}\right].
    f_{\textsc{v}}(\sigma_i - \sigma_j)=\sum_{m=-\infty}^{\infty} e^{-1/2 \beta[2\pi/q (\sigma_i - \sigma_j)-2 \pi m]^{2}}.
\end{equation}
The model has the nice property that it is self-dual on the square lattice for all integers $q$ in contrast to the original clock model. 
The duality relation is written as $\beta^* = q^2/4\pi^2\beta$ and the self-dual point is given by 
$\beta_{\rm sd} = q/2\pi$.

The renormalization group flow equations of the Villain model are written as \cite{elitzurPhaseStructureDiscrete1979}
\begin{subequations}\label{eq:RG_flow}
	\renewcommand{\theequation}{\theparentequation.\arabic{equation}}
    \begin{align} 
        \frac{dx}{d\ln b} &= \frac{q^2}{4}y_q^2 - x^2 y^2, \\
        \frac{dy}{d \ln b} &= (2-x)y, \\
        \frac{dy_q}{d\ln b} &= (2-\frac{q^2}{4x})y_q,
    \end{align}
\end{subequations}
where $b$ is the rescaling factor, and the parameters $x$, $y$, and $y_q$ are initially defined by
\begin{equation}\label{eq:init}
    \begin{cases}
    x = \pi\beta,\\
    y = 2\pi \exp(-\pi^2\beta/2), \\
    y_q = 2\pi \exp(-q^2/8\beta).
    \end{cases}
\end{equation}

The parameter $x$ corresponds to the effective inverse temperature of the system under renormalization.
The parameter $y$ characterizes the effect of vortices on spin configurations. When $y$ is a relevant operator, 
i.e., $x^\textsc{r} < 2$, where the superscript $\textsc{r}$ stands for renormalized, the abundance of vortices destroys the order of spins, driving the system to 
the infinite-temperature fixed point for the high-T disordered phase.
In contrast, the parameter $y_q$ characterizes the effect of discretization of clock spins on spin-wave excitations.
When $y_q$ is a relevant operator, i.e., $x^\textsc{r} > q^2/8$, the temperature is too low to sufficiently excite the discretized spin-wave excitation so that the clock spins tend to point in one of the $q$ directions
and the renormalization flows are driven to the zero-temperature fixed point.
The QLRO phase emerges when both $y^{\textsc{r}}$ and $y^{\textsc{r}}_q$ are irrelevant, i.e., $y^\textsc{r} = y^\textsc{r}_q = 0$, which corresponds to a line of fixed points 
with $2 \leq x^\textsc{r} \leq q^2/8$. The two-point correlation function in this phase diverges as 
\begin{equation}
    g(\textbf{r}) \sim \frac{1}{\|\textbf{r}\|^{1/2x^\textsc{r}}} = \frac{1}{\|\textbf{r}\|^{\eta}}\quad(\|\textbf{r}\|\to \infty),
\end{equation}
from which we know the anomalous dimension $\eta = 1/2x^\textsc{r}$.
The boundaries of the middle phase correspond to the two critical points for $q \geq 5$,  which allows us to compute $\eta$ at these two special points 
\begin{equation}\label{eq:Villain_eta}
\begin{aligned}
    x^\textsc{r}&=2, &&\eta = 1/4\quad \ \,\text{for} \ \beta =\bco\\
    x^\textsc{r}&=q^2/8, &&\eta= 4/q^2\quad \text{for}\ \beta = \bct. \\
    % \beta &= \beta_{\rm sd}: \quad x^\textsc{r}=q/2, &&\eta(\beta_{\rm sd}) = 1/2x^\textsc{r} = 1/q.
\end{aligned}
\end{equation}

Moreover, at the self-dual point $\beta_{\rm sd} = q/2\pi$, one may find $x = q/2$ and $y = y_q=2\pi \exp(-\pi q/4)$ 
by Eq.~\eqref{eq:init}. According to Eqs.~\eqref{eq:RG_flow}, under the process of the renormalization, 
the parameters $y$ and $y_q$ remain identical, leaving $x$ invariant with $x^{\textsc{r}} = q/2$, 
which gives $\eta(\beta_{\rm sd}) = 1/q$.

\section{Algorithm and Sampled quantities}
\label{Sec: Algorithm}
For self-completeness and convenience of readers, we describe the worm algorithm for the clock model in detail 
and define the quantities sampled during the simulation.

\subsection{Worm Algorithm}
\label{subsec: Worm Algorithm}

The worm algorithm works on an extended configuration space composed of the original partition function space (the $\scrZ$ space) and the two-point correlation function space (the $\scrG$ space).
The $\scrZ$ space contains the set of closed flow configurations [see Fig.~\ref{fig:HT-flows}(a)]. In contrast, the $\scrG$ space consists of open flow configurations, which have two defects $\scrI, \scrM$ violating the $q$-modular flow conservation, as illustrated in Fig.~\ref{fig:HT-flows}(b).
For the high-$T$ expansion, the $\scrG$ space describes the two-point correlation in the original spin representation. 
Taking any two defects $\scrI$ and $\scrM\in V$, we define the quantity $G_\textsc{h}(\scrI, \scrM) = \langle \mathbf{S}_\scrI \cdot \mathbf{S}_\scrM \rangle \scrZ$, where $\scrZ$ is the partition function.
Then, by performing similar manipulations used in deriving Eq.~\eqref{eq:HT_expansion}, $G_\textsc{h}(\scrI, \scrM)$ can be expressed in terms of the high-$T$ flows:
	\begin{align}
		&G_\textsc{h}(\scrI, \scrM) = \langle \mathbf{S}_\scrI \cdot \mathbf{S}_\scrM \rangle \scrZ \nonumber\\
		&= \sum_{\{\sigma\}}\cos (\frac{2\pi}{q} (\sigma_\scrM - \sigma_\scrI)) \nonumber\\
            &\times \exp\left[\beta \sum_{\langle ij \rangle}\cos(\frac{2\pi}{q}(\sigma_i - \sigma_j))\right]  \nonumber\\
		% e^{\beta \sum_{\langle ij \rangle}\cos(\frac{2\pi}{q}(\sigma_i - \sigma_j))} \nonumber\\
		&= q^{|V|} \sum_{\{N\}: \nabla \cdot \mathbf{N} = S_{\scrI\scrM}} \left(\prod_{\langle ij \rangle\in E}F(N_{ij})\right).
	\end{align}
The constraint $\nabla \cdot \mathbf{N} = S_{\scrI\scrM}$ is shorthand for $(\nabla \cdot \mathbf{N})_i \bmod q= S_{\scrI\scrM} (i)$ for $i \in V$ and $S_{\scrI\scrM}$ is a function of sites:
\begin{equation}\label{eq:source}
	S_{\scrI \scrM}(i) =
	\begin{cases}
		1 & \text{for}\  i = \scrI\\
		q-1 & \text{for}\  i = \scrM\\
		0 & \text{otherwise}.
	\end{cases}
\end{equation}
Different from the previous divergence-free condition, Eq.~\eqref{eq:source} requires that the flows originate at $\scrI$ with divergence 1 and end at $\scrM$ with divergence $q-1$.
%($-1$ is equivalent to $q-1$ in the modulo $q$ sense).
The partition function of the $\scrG$ space is then the summation with $\scrI,\scrM$ $(\scrI\neq \scrM)$ varied,
\begin{equation}\label{eq:G_space}
	\scrG = \sum_{\scrI \neq \scrM} G_\textsc{h}(\scrI, \scrM) = q^{|V|} \sum_{\{N\} \in \scrS} \left(\prod_{\langle ij \rangle\in E}F(N_{ij})\right),
\end{equation}
where $\scrS$ is the set of open configurations that have one defect with divergence $1$ and one defect with divergence $q-1$.
As we can see, the $\scrG$ space is tailored to study the two-point correlation and susceptibility since the latter is the summation of the former over all possible coordinates of $(\scrI, \scrM)$. 
In principle, the function $S_{\scrI \scrM}$ can be modified to take other values, which can be used to study other forms of the two-point correlation function.
Here we restrict ourselves to Eq.~\eqref{eq:source} because it provides a convenient way to sample the susceptibility $\chi_\textsc{h}$, as explained below.
% which is the quantity we need in studying the phase transitions of the model.

Since the weight functions $F(N_{ij})$ in Eqs.~\eqref{eq:HT_expansion} and \eqref{eq:G_space} are identical, the $\scrZ$ space and $\scrG$ space can be combined to form
a larger configuration space, whose partition is written as
\begin{equation}
	\scrZ_{\text{ext}} = C \scrZ + \scrG.
\end{equation}
Here $C$ is the parameter of the algorithm which controls the switching probability between $\scrZ$ space and $\scrG$ space. We set $C = L^d$ $(d=2)$ throughout our work and get
% 	\begin{equation} 
		\begin{align}\label{eq:susceptibility}
			\scrZ_{\text{ext}} &= L^d \scrZ + \scrG = \sum_{\scrI, \scrM \in V} G_\textsc{h}(\scrI, \scrM)\nonumber \\
			&= \left\langle \left(\sum_{i \in V} \mathbf{S}_i\right)^2 \right\rangle \scrZ = \chi_\textsc{h} L^d \scrZ,
		\end{align}
% 	\end{equation}
where $\chi_\textsc{h}$ is the magnetic susceptibility in the original spin representation and we have used $\scrZ = \langle \mathbf{S}_{\scrI}^2 \rangle \scrZ = G(\scrI, \scrI)$. From Eq.~\eqref{eq:susceptibility}
we know that $\chih$ can be expressed as the ratio between $\scrZ_{\text{ext}}$ and $L^d \scrZ$ of the high-$T$ flows, i.e., the steps between the two consecutive events of hitting the $\scrZ$ space in the case of $C = L^d$.

Because the low-$T$ expansion has a form similar to the high-$T$ expansion, we can construct analogous
quantities out of the low-$T$ flows, which should exhibit dual scaling behaviors.
For defect $\scrI, \scrM \in V^*$, we define $G_\textsc{l}(\scrI, \scrM)$ as
\begin{equation}
	G_\textsc{l}(\scrI, \scrM) = q \sum_{\{N^*\}: \nabla \cdot \mathbf{N}^* = S_{\scrI\scrM}} \left(\prod_{\langle ij \rangle \in E^*}F^*(N^*_{ij})\right),
\end{equation}
where the function $S_{\scrI\scrM}(i)$ is the same as Eq.~\eqref{eq:source} but with $i$, $\scrI$, and $\scrM$ now vertices of the dual lattice $G^*$.
The corresponding susceptibility-like quantity $\chi_\textsc{l}$ from the low-$T$ flows is defined by
\begin{equation}
	\chi_\textsc{l} = \frac{1}{L^d}\sum_{\scrI, \scrM \in V^*} \frac{G_\textsc{l} (\scrI, \scrM)}{\scrZ}.
\end{equation}
Moreover, since the square lattice with periodic boundary conditions is a self-dual lattice, we can simulate the low-$T$ flows using the same algorithm by only replacing $F(N)$ with $F^*(N)$ without further modification to the underlying lattice.

We now describe the worm algorithm for the high-$T$ flows in detail. The same procedure applies to the low-$T$ flows. The algorithm samples configurations in the extended space $\scrZ_{\text{ext}}$. A state in $\scrZ_{\text{ext}}$ can be identified by its bond configuration together with the positions of $\scrI$ and $\scrM$. Therefore, to sample configurations in $\scrZ_{\text{ext}}$, 
one can move defect $\scrI$ or $\scrM$ locally and update the bond configuration accordingly to keep it a valid open configuration. The Metropolis criterion is used to decide whether this move is accepted or not. More specifically, the basic procedure of algorithm is as follows:

\begin{enumerate}
	\item If $\scrI = \scrM$, choose a new site $\scrI^\prime \in V$ randomly with probability $1/L^d$ and set $\scrI = \scrM = \scrI^\prime$. If $\scrI \neq \scrM$, start from step 2.
	\item Starting from a configuration $\mu$, randomly pick a defect with equal probability, say $\scrI$.
	% \item Randomly choose a nearest neighbour $P$ of the chosen defect $I$ (or $M$). We propose the increased update $N_{IP}\rightarrow N_{IP}^\prime = (N_{IP}+1)\bmod q$ (or $N_{MP} \rightarrow N_{MP}^\prime = (N_{MP}-1)\bmod q$ ) and  accept it with probability $P_{\text{acc}}^{+}$ and $P_{\text{acc}}^{-}$ respectively.
	\item Randomly choose a nearest neighbor $\scrI^\prime$ of the chosen defect $\scrI$. We propose the update $N_{\scrI \scrI^\prime}\to N_{\scrI \scrI^\prime}^\prime \coloneqq [N_{\scrI \scrI^\prime}-\text{sgn}(\scrI \to \scrI^\prime)]\bmod q$ (replace $-$ with $+$ for the move of defect $\scrM$) to get a new configuration $\nu$ and accept it with probability $P^{\text{acc}}_{\mu \to \nu}$.
	\item If the proposal is accepted, assign $\scrI^\prime$ to be the new defect $\scrI \coloneqq \scrI^\prime$.
\end{enumerate}

The acceptance probability $P^{\text{acc}}_{\mu \to \nu}$ is calculated according to the Metropolis-Hastings scheme
\begin{equation}
   P^{\text{acc}}_{\mu \to \nu} = \min\left\{1, \frac{\scrA_{\mu \to \nu} W_{\nu} }{ \scrA_{\nu\to \mu} W_{\mu} } \right\},
\end{equation}
where $\scrA_{\mu \to \nu}$ ($\scrA_{\nu \to \mu}$) is the proposal probability with which we propose the update from configuration $\mu$ to $\nu$ ($\nu$ to $\mu$),
and $W_{\mu}$ ($W_{\nu}$) is the statistical weight of $\mu$ ($\nu$). According to the sectors $\mu$ and $\nu$ belong to, the ratio $\scrA_{\mu \to \nu}/ \scrA_{\nu\to \mu}$ can take three possible values
\begin{equation}
	\frac{\scrA_{\mu \to \nu}}{ \scrA_{\nu\to \mu}} =
	\begin{cases}
		1,   & \mu, \nu \in \scrG \\
		L^d,   & \mu \in \scrG, \nu \in \scrZ \\
		1/L^d, & \mu \in \scrZ, \nu \in \scrG.
	\end{cases}
\end{equation}
Thus, $P^{\text{acc}}_{\mu \to \nu}$ generally has different expressions for each case. Fortunately, with the choice of the relative weight $C =L^d$, 
$P^{\text{acc}}_{\mu \to \nu}$ reduces to one formula for the three cases, independent of the sectors $\mu$ and $\nu$ belong to. To be more specific, 
we have $P^{\text{acc}} = \min\left\{1, F(N^\prime_{\scrI \scrI^\prime})/ F(N_{\scrI \scrI^\prime}) \right\}$ for the update $N_{\scrI \scrI^\prime}\to N^\prime_{\scrI \scrI^\prime}$;
a similar expression can be derived for moving defect $\scrM$.
To simulate the low-$T$ flows,  we only need to replace $F(N)$ with $F^*(N^*)$ and proceed analogously.

\subsection{Sampled quantities}
\label{Subsec: Sampled quantites}
In this work we use the worm algorithm to simulate the $q$-state clock model on the square lattice ($L\times L$) with periodic boundary conditions in its two flow representations. Since our aims are to determine
the critical points and demonstrate the duality of the model, it suffices to measure the following quantities in the partition function space
$\scrZ$ or in the extended space $\scrZ_{\rm ext}$.

\begin{enumerate}[(i)]
	\item First is the worm-returning time $\scrT_\textsc{h}$ in the high-$T$ flows and $\scrT_\textsc{l}$ in the low-$T$ flows. 
	We define a worm cycle to be the Markov chain between the two consecutive events of the worm configuration hitting
	the $\scrZ$ space, i.e., defects $\scrI$ and $\scrM$ coincide. Then for each worm cycle, we define the returning time 
	as the number of update steps consisting of the cycle. Thus, this quantity is only measured when configuration $\mu$ is in the $\scrZ$ space.
	\item Second is the Euclidean distance of the two defects $\scrL_{\scrI \scrM}$ in the high-$T$ flows, which is sampled in the extended space $\scrZ_{\text{ext}}$ after each MC sweep.

\end{enumerate}
The corresponding ensemble average is taken as (a) the susceptibility $\chi_\textsc{h} = \langle \scrT_\textsc{h}  \rangle $ in the high-$T$ flows 
and the susceptibility $\chi_\textsc{l} = \langle \scrT_\textsc{l}  \rangle$ in the low-$T$ flows, (b) the difference between the two susceptibilities $\chi_{\rm diff} = \langle \scrT_\textsc{h}\rangle - \langle\scrT_\textsc{l} \rangle$ at the same temperature, and (c) the correlation length $\xi = \langle \scrL_{\scrI \scrM} \rangle$, which is defined as $\xi = \int  \|\mathbf{r}\| g(\mathbf{r}) d\mathbf{r}\Big/\int g(\mathbf{r}) d\mathbf{r}$.
% Here we denote the sampling of the unwrapped distance $\scrU_{\rm IM}$ is in $Z_{\rm ext}$ space uniformly rather than in the $\scrZ$ space.

\section{Results}
\label{Sec: Results}
In this section we provide numerical results that give the estimates of the critical points $\bco$ and $\bct$ and explore the duality in the $q$-state clock model for $q \geq 5$.
We perform least-squares fits of our Monte Carlo data to the expected \textit{Ansatz}. As a precaution against correction-to-scaling terms that we miss including in the fitting \textit{Ansatz}, we impose a lower cutoff $L \ge L_{\rm m}$ on the data points admitted in the fits. We systematically study the effect on the residuals (denoted by $\rm{chi}^2$) by increasing $L_{\rm m}$. In general, the preferred fit for any given \textit{Ansatz} corresponds to the smallest $L_{\rm m}$ for which the goodness of the fit is reasonable and for which subsequent increases in $L_{\rm m}$ do not cause the $\chi^2$ value to drop by vastly more than one unit per degree of freedom $\mathcal{D}$. In practice,
by “reasonable” we mean that $\chi^2/ \mathcal{D} \approx 1$. The systematic error is obtained by comparing estimates from various reasonable fitting \textit{Ans{\"a}tze}. \par 

\subsection{Estimation of $\bco$ and $\bct$}
\label{Subsec: critical_points}
\begin{figure}[t]
	% \centering
	\includegraphics[width=.71\columnwidth]{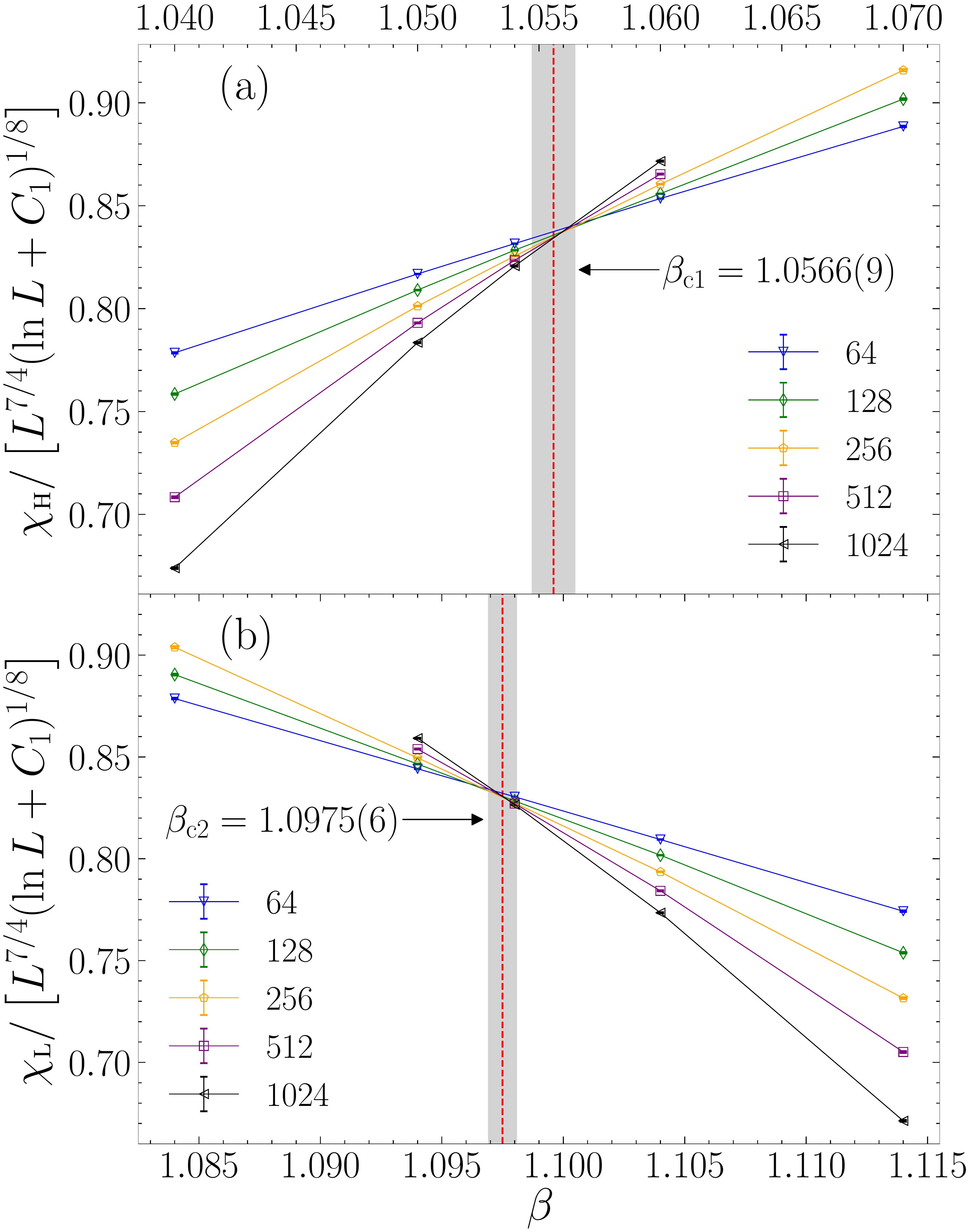}
	\caption{Scaled susceptibility (a) $\chi_\textsc{h}(\beta, L)/L^{7/4}(\ln L+C_1)^{1/8}$ and (b) $\chi_\textsc{l}(\beta, L)/L^{7/4}(\ln L+C_1)^{1/8}$
	versus the inverse temperature $\beta$ for $q=5$. The constant $C_1$ is set to (a) 3.6 and (b) 4.0. 
	The vertical red dashed line presents the central value of our estimate and the shadow shows the error bar.}
	\label{fig:chiHL_q5}
\end{figure}

We use observables $\chih$ and $\chil$ to estimate $\bco$ and $\bct$, respectively.
As we explained in Sec.~\ref{Sec: Algorithm}, the ensemble average of the worm returning time in the high-$T$ flows $\chih$
corresponds to the magnetic susceptibility of the system, 
which has the following scaling form at the high-$T$ transition point $\bco$
\begin{equation}\label{eq:chih_scaling}
	\chih(\bco) \sim \int_{r < \xi} g(\mathbf{r}) d^2 r\sim \xi^{2-\eta}(\ln\xi)^{-1/8},
\end{equation}
where $\xi$ is the correlation length and $\eta = 1/4$. 
It is also noted that, in Eq.~\eqref{eq:chih_scaling}, there is an extra logarithmic dependence of $\chih(\bco)$,
which originates from the multiplicative logarithmic correction to the correlation function $g(\mathbf{r})$ at $\bco$ \cite{elitzurPhaseStructureDiscrete1979} similar to the $XY$ model \cite{kosterlitzCriticalPropertiesTwodimensional1974, pelissettoRenormalizationgroupFlowAsymptotic2013, wangPercolationTwodimensionalModel2021}.
As for finite systems, $\xi$ is cut off by the linear system size.
% as $\xi = \alpha L$, with $\alpha$ a nonuniversal constant.
Using the linear system size $L$, we then have $\chih(\bco, L) \sim L^{7/4}(\ln L + C_1)^{-1/8}$, where $C_1$ introduces a characteristic length scale for the multiplicative logarithmic correction.
In Fig.~\ref{fig:chiHL_q5}(a) we plot the scaled susceptibility $\tilde{\chi}_\textsc{h}(\beta, L) = \chi_\textsc{h}(\beta, L)/ L^{7/4}(\ln L+C_1)^{1/8}$
versus the inverse temperature $\beta$ using our MC data for $q=5$. As it is shown, there is an excellent intersection for different system sizes at $\bco$, 
which confirms the scaling form~\eqref{eq:chih_scaling} of $\chih$.

For observable $\chil$, on the other hand, it neither corresponds to the magnetic susceptibility of original spins nor has known analytical results about their
critical behaviors. Nevertheless, based on our previous duality argument in Sec.~\ref{subsec:self-duality},
we expect $\chil$ at the low-$T$ transition point $\bct$ to have the same scaling behavior as that of $\chih$ at the high-$T$ transition point $\bco$.
This expectation is supported by Fig.~\ref{fig:chiHL_q5}(b), as an excellent intersection point is present for the scaled susceptibility $\tilde{\chi}_{\textsc{l}}(\beta, L) = \chil(\beta, L)/ L^{7/4} (\ln L + C_1)^{1/8}$ with various system sizes.

\begin{figure}[t]
	\centering
	\includegraphics[width=\columnwidth]{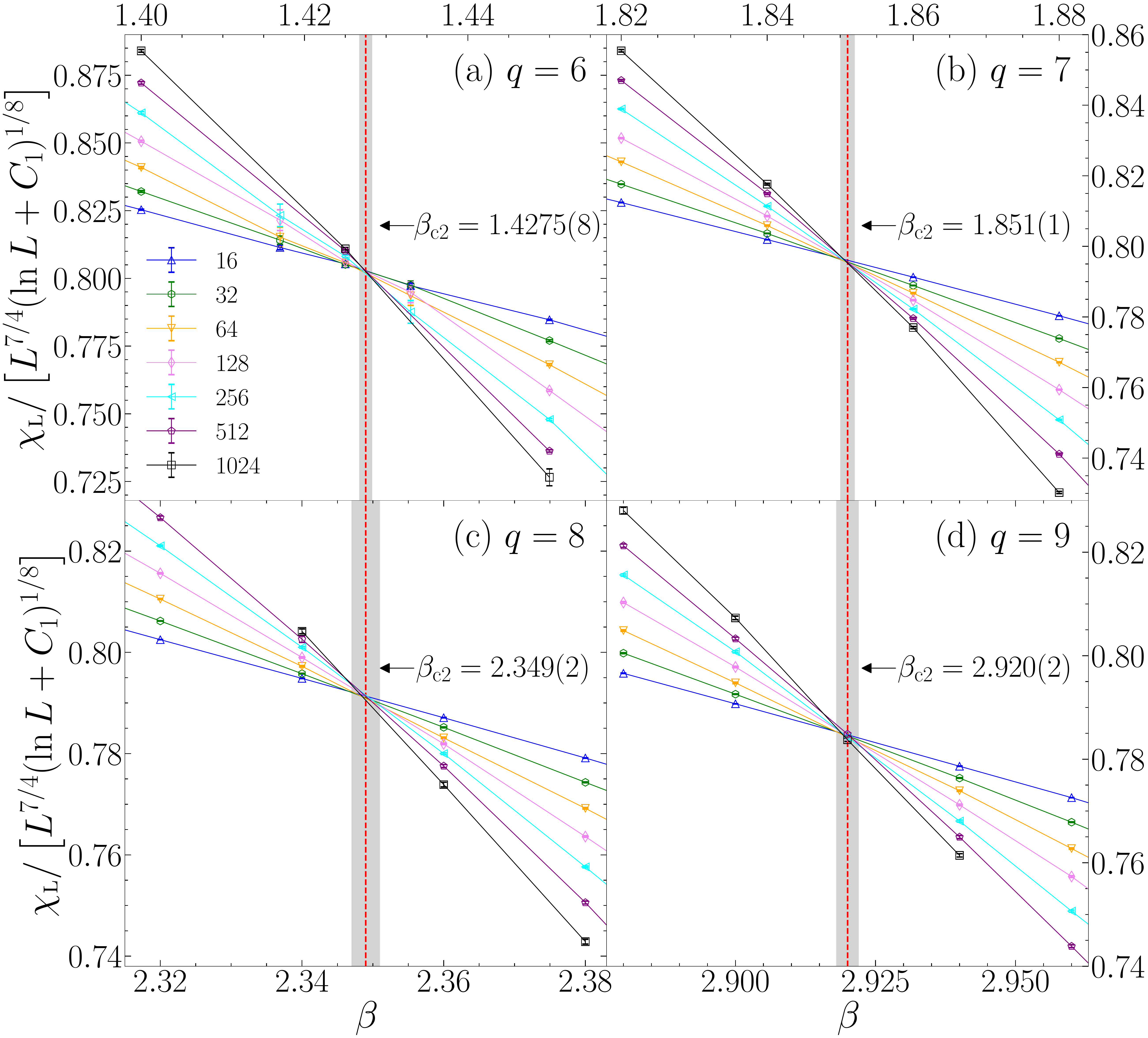}
	\caption{Scaled susceptibility $\chi_\textsc{l}(\beta, L)/ L^{7/4}(\ln L+C_1)^{1/8}$ versus the inverse temperature $\beta$ for (a) $q=6$, (b) $q=7$, (c) $q=8$, and (d) $q=9$.
	The constants $C_1$ are set to 4.9, 4.9, 5.0, and 5.3, respectively. The vertical red line represents the 
	central value of $\bct$, and the shadow shows the error bar.}
	\label{fig:chiL6_to_9}
\end{figure}

We then estimate $\beta_{\rm c1}$ and $\beta_{\rm c2}$ systematically. Instead of using $\chih$ to determine $\bct$, here we choose $\chil$ to estimate $\bct$ because it suffers weaker finite-size corrections around $\bct$, as illustrated in Figs.~\ref{fig:chiHL_q5}(b) and \ref{fig:chiL6_to_9}.
We perform the least-squares fits to $\chih(\beta, L)$ and $\chil(\beta, L)$ via the finite-size-scaling \textit{Ansatz}
% \begin{equation}
	\begin{align}\label{eq:fitting_ansatz_beta_c}
		\chi&(\beta,L) = L^{7/4}(\ln L + C_1)^{1/8} \Big[a_0 + \sum_{k=1}^3 a_k\epsilon^k(\ln L +C_2)^{2k}\nonumber \\
		&+ d_1 L^{y_1} + d_2 L^{y_2}+n_0\epsilon+n_1 \epsilon^2(\ln L +C_2)^{2}\Big].
	\end{align}
% \end{equation}
Here $\epsilon$ stands for  $\bco - \beta$ when $\chi$ represents $\chih$ and for $\bct - \beta$ when $\chi$ represents $\chil$.
The term $a_{k}$ on the right-hand side of Eq.~\eqref{eq:fitting_ansatz_beta_c} come from the Taylor expansion of the scaled susceptibility with respect to the scaling field $\epsilon (\ln L/L_0)^2 = \epsilon(\ln L + C_2)^2$ around the critical point. 
The terms $d_1$ and $d_2$ account for the additive finite-size corrections with $y_2 < y_1 <0$.
The $n_0$ term describes the asymmetry dependence of the scaling function. The $n_1$ term originates from the nonlinearity of the RG invariant function 
$b_1 \epsilon + b_2 \epsilon^2 + \cdots$ as the scaling field \cite{pelissettoRenormalizationgroupFlowAsymptotic2013}.

\begin{table*}[t]
\caption{Fitting results of $\bco$ and $\bct$ for $q=5$ from $\chi_\textsc{h}$ and $\chil$ with \textit{Ansatz}~\eqref{eq:fitting_ansatz_beta_c}.
The column of $\beta_{\rm c}$ corresponds to $\bco$ when $\mathcal{O}$ is taken as $\chih$ and corresponds to $\bct$ when $\mathcal{O}$ is taken as $\chil$.
}
\label{tab:chiHL5}
\centering
\setlength\tabcolsep{1.3pt}
\begin{tabular}{|l|lclclllllcllc|}
\hline
$\mathcal{O}$ &$L_{\rm m}$  	&$\beta_{\rm c}$ 	&\multicolumn{1}{c}{$C_1$} 	&\multicolumn{1}{c}{$C_2$} 	&\quad$a_0$ 	&\multicolumn{1}{c}{$a_1$}	&\multicolumn{1}{c}{$a_2$} 	
&\multicolumn{1}{c}{$a_3$} 	&\multicolumn{1}{c}{$n_0$} 	&\multicolumn{1}{c}{$n_1$}   &\multicolumn{1}{c}{$d_1$} 
&\multicolumn{1}{c}{$d_2$}  &\multicolumn{1}{c|}{${\chi}^2/{\mathcal{D}}$} 	\\
\hline
\multirow{5}{*}{$\chih$}
&24    &1.055\,7(5) 	&~3(1)   	&0.1(2)    	&0.84(1)   	&~-0.171(9) 	&-0.005(8) 	&-0.038(9) 	&-0.6(1)   	&-0(3)     	&0.3(1)    	&-0.6(8) &21.8/19\\ 
&32    &1.055\,6(9) 	&~3(2)      	&0.1(3)    	&0.84(3)   	&~-0.17(1)  	&-0.008(9) 	&-0.04(1)  	&-0.7(2)   	&0(4)      	&0.3(3)    	&-0(2)   &20.7/15\\ 
&24    &1.055\,3(2) 	&~3.6(3)    	&- &0.832(5)  	&~-0.173(2) 	&-0.009(5) 	&-0.040(5) 	&-0.68(2)  	   &\multicolumn{1}{c}{-}	&0.24(2)   	   	&\multicolumn{1}{c}{-} 		&22.5/22\\ 
&32    &1.055\,4(2) 	&~3.5(4)    	&- &0.835(6)  	&~-0.173(2) 	&-0.009(6) 	&-0.040(5) 	&-0.70(3)  	   &\multicolumn{1}{c}{-}	&0.25(3)   	   	&\multicolumn{1}{c}{-} 		&20.9/18\\ 
&48    &1.055\,5(3) 	&~3.3(6)    	&- &0.838(9)  	&~-0.173(2) 	&-0.005(7) 	&-0.043(6) 	&-0.73(5)  	   &\multicolumn{1}{c}{-}	&0.27(6)   	   	&\multicolumn{1}{c}{-} 		&18.3/14\\ 
\hline
\multirow{4}{*}{$\chil$}
&32   &1.097\,5(3) 	&~3.4(5)    	&0.0(2)    	&0.836(8)  	&~~0.157(7)  	&-0.029(8) 	&~0.016(6)  	&~0.7(1)    	&0.4(1)    	&0.24(3)   	&\multicolumn{1}{c}{-} 	&27.0/18\\ 
&48   &1.097\,6(5) 	&~4(1)      	&-0.3(3)   	&0.83(1)   	&~~0.17(1)   	&-0.04(1)  	&~0.017(9)  	&~1.0(2)    	&0.4(2)    	&0.22(8)   	&\multicolumn{1}{c}{-} &20.5/14\\ 
&32    &1.097\,5(3) 	&~3.5(5)  &\multicolumn{1}{c}{-} 	&0.836(7)  	&~~0.1587(10)	&-0.030(8) 	&~0.016(6)  	&~0.74(3)   	&0.4(1)    	&0.23(3)   &\multicolumn{1}{c}{-}		&27.1/19\\ 
&48    &1.097\,3(4) 	&~3.1(8)  &\multicolumn{1}{c}{-}  	&0.84(1)   	&~~0.157(1)  	&-0.034(9) 	&~0.014(6)  	&~0.80(5)   	&0.5(2)    	&0.27(7)   &\multicolumn{1}{c}{-}		&21.5/15\\ 
\hline
\end{tabular}
% }
\end{table*}

For convenience, we simply set $y_1=-1$ and $y_2=-2$.  Table~\ref{tab:chiHL5} reports the fitting results of $q=5$, where parameters set to 0 are denoted by `--'.
In the case of $\chi_{\textsc{h}}$, we first leave all parameters free, which gives the estimate $\bco = 1.0556(9)$,
and we find that $C_2, n_1,$ and $d_2$ are consistent with 0. Then we set $C_2 = n_1 = d_2=0$ and get $\bco = 1.0554(4)$.  
By comparing estimates from various ansatz, we finally obtain $\bco = 1.0556(9)$. Similarly, in the fit of 
$\chil$, we first leave $C_2$ and $d_2$ free and find that both of them are consistent with zero. We then perform
another fit with $C_2 = d_2 = 0$. Both fits give the stable estimate $\bct = 1.097\,5(6)$, which agrees with the recent MC result \cite{surunganBerezinskiiKosterlitzThouless2019}.
A similar analysis is applied to determine the critical points $\bco$ and $\bct$ for other values of $q$, and the details of the fitting are presented in the Appendix. 
%summarized in Tables~\ref{tab: chiH69} and \ref{tab: chiL6_to_9}.
In Fig.~\ref{fig:chiL6_to_9}, we plot the scaled susceptibility $\tilde{\chi}_{\textsc{l}}$ versus $\beta$ for $q=6,7,8,9$ and indicate our estimates by the vertical red lines.

The final results of $\bco$ and $\bct$ are summarized in Table~\ref{tab: summary}. Our estimates of the critical points are consistent
with the previous MC results with the precision being significantly improved, as shown in Table~\ref{table: different works}.
On the other hand, apart from the inconsistency among the TN results, some of them are nearly excluded by our estimates if the quoted error margins are taken seriously into account. 
In Fig.~\ref{fig: SDKc} we plot our estimates of $\bco$ and $\bct$ as a function of $q$. 
As Fig.~\ref{fig: SDKc} shows, one may find $\bco \sim O(1)$,  which quickly converges to the 2D $XY$ model transition point $\beta_{\rm BKT} = 1.119\,96(6)$ \cite{wangPercolationTwodimensionalModel2021} as $q$ increases,
and $\bct\sim O(q^2)$, consistent with the statement in Ref.~\cite{ortizDualitiesPhaseDiagram2012}.
Further, the least-squares fit of $\bct$ with the formula $\bct(q) = a_0 + a_1q + a_2q^2$ gives $a_0 = 0.65(6), a_1 = -0.11(2),$ and $a_2=0.041(1)$. Comparing with the conjectured formula $\beta_{\rm c2}= q^2/A + B q + C + D e^{-\frac{\pi^2q^2}{A}}$  based on the Villain clock model from Ref.~\cite{borisenko2012phase}, we find that both of them give the same leading behavior, i.e., the estimate $a_2=0.041(1)$ agrees with the inverse of $A=25.89(115)$, both of which are consistent with $1/8\pi$.

\subsection{Duality between $\bco$ and $\bct$}
\label{subsec:critical_points_duality}

In general, models in the same universality class are governed by the same fixed point and 
share the same asymptotic phenomena, such as the critical exponents and amplitude ratios, etc.
% have the same universal values, like the critical exponents, Binder ratios and amplitude ratios, etc.} 
Hence, as illustrated in Sec.~\ref{Subsec: critical_points}, the high-$T$ flows at $\bco$ and low-$T$ flows at $\bct$ belong to the same universality class.
Moreover, it is interesting to note that, for $q=5$, the nonuniversal parameters $a_0$ and $C_1$ of $\chih$ and $\chil$ in Table~\ref{tab:chiHL5} are numerically consistent with each other, 
which suggests the more stringent duality relation $\chih(\bco, L) = \chil(\bct, L)$. For larger $q$, this relation
seems to hold approximately. From Table~\ref{tab: chiHL_appendix} in the Appendix, it is also noted that the values of $a_0$ and $C_1$ for $\chih(\bco, L)$ and $\chil(\bct, L)$ are nearly independent of $q$, even though the critical point $\bct(q)$ has a clear dependence on $q$, which
implies that these models are not only governed by the same fixed point but also close to each other in the critical surface in the RG analysis.

\begin{figure}[htp]
	\centering
	\includegraphics[width=.85\columnwidth]{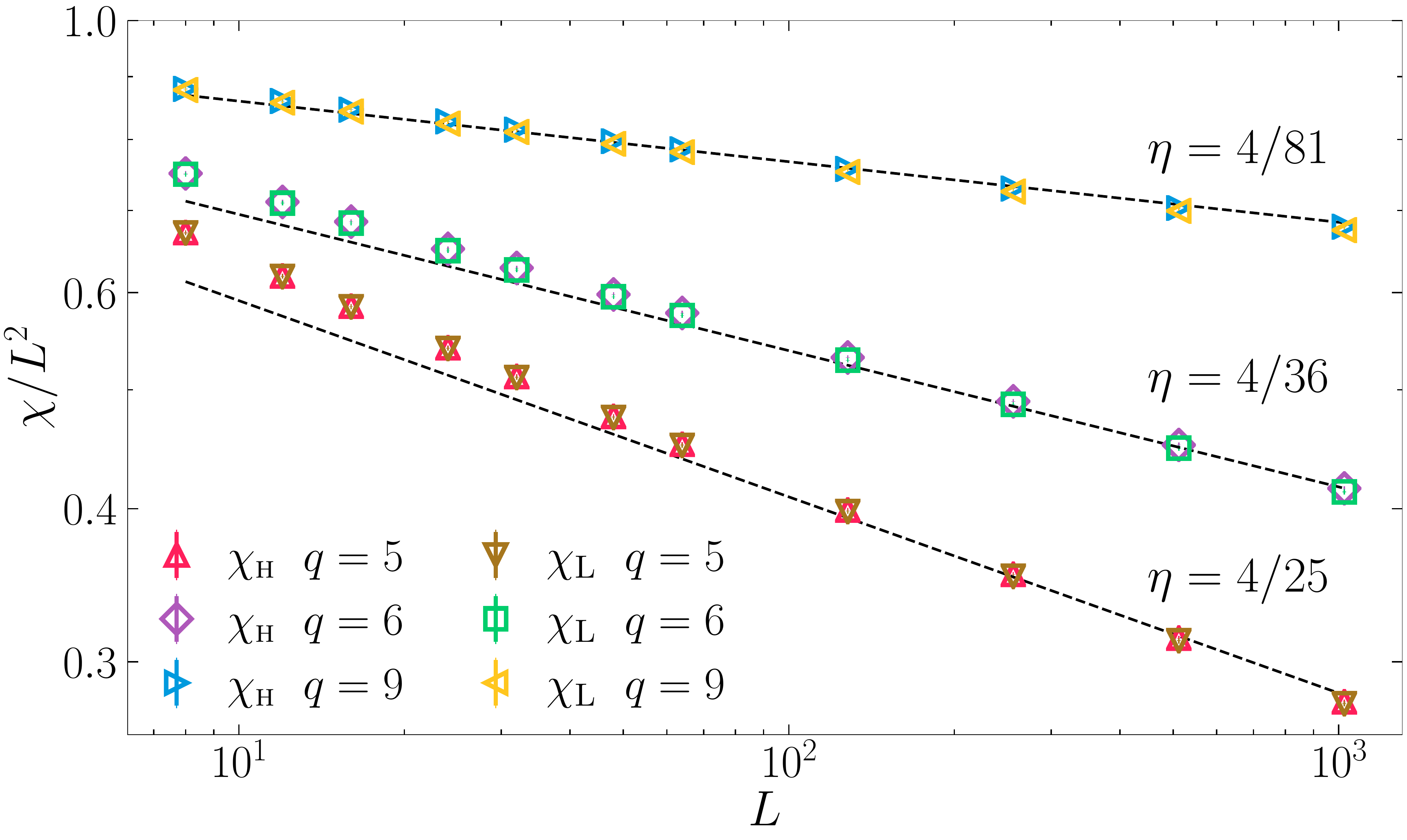}
	\caption{Scaled susceptibilities $\chi_\textsc{l}(\bco, L)/L^{2}$ and $\chi_\textsc{h}(\bct, L)/L^{2}$ versus $L$ on a log-log scale. 
	The dashed line represents $\chi(L)/L^2= a_0 L^{-\eta}$ with $\eta=4/q^2$ for $q=5,6,9$. For each $q$, the data points of $\chih(\bct, L)$ and $\chil(\bco, L)$
	are very close to each other, which indicates that they not only share the same scaling behavior but also have the same amplitude.
	}
	% There may be a multiplicative logarithmic correction causing the deviations of the data points from the dashed lines.}
	\label{fig:eta}
\end{figure}
To further demonstrate the duality between $\bco$ and $\bct$,  we analyze the data of $\chi_\textsc{l} (\beta_{\rm c1})$ and $\chi_\textsc{h}(\beta_{\rm c2})$. 
In Sec.~\ref{Subsec: critical_points} we make use of the duality between $\{N\}$ at $\bco$ and $\{N^*\}$ at $\bct$ to infer the scaling of $\chil(\bct, L)$.
If the duality between $\bco$ and $\bct$ is preserved for $q > 5$, there should also exist a connection between $\{N\}$ at $\bct$ and $\{N^*\}$ at $\bco$.
As shown in Fig.~\ref{fig:eta}, irrespective of the value of $q$,  $\chih(\bct, L)$ and $\chil(\bco, L)$ are nearly identical even for a linear system size as small as $L=8$, 
vividly illustrating the duality relation between the two critical points and between the two flow representations.
We numerically determine the exponents $\eta_{\textsc{l}}(\bco)$ for the low-$T$ flows and $\eta_{\textsc{h}}(\bct)$ for the high-$T$ flows 
by fitting the MC data with the \textit{Ansatz}

\begin{equation}\label{eq:fitting_ansatz_eta}
	% \chi^{\rm{L}/\rm{H}} (L)L^{-2} = L^{-\eta} (a_0 + d_1 L^{y_1}).
	\chi(L)/L^{2} = L^{-\eta} (a_0 + d_1 L^{y_1}),
\end{equation}
where $\chi(L)$ stands for $\chi_{\textsc{l}}(\bco, L)$ or $\chi_{\textsc{h}}(\bct, L)$. The $d_1$ term accounts for additive corrections.
The final results are $\eta_{\textsc{l}}(\bco) =$ 0.169(3), 0.116(2), 0.052\,3(7) and $\eta_{\textsc{h}}(\bct)=$0.168(2), 0.116(1), 0.051(2) for $q=5,6,9$.
There is a minor deviation between the estimates and the expected value of $4/q^2$, which is probably caused by logarithmic corrections.

\subsection{Self-dual point $\beta_{\rm sd}$ for $q > 5$}
\begin{figure}[b]
    \centering
    \includegraphics[width=.85\columnwidth]{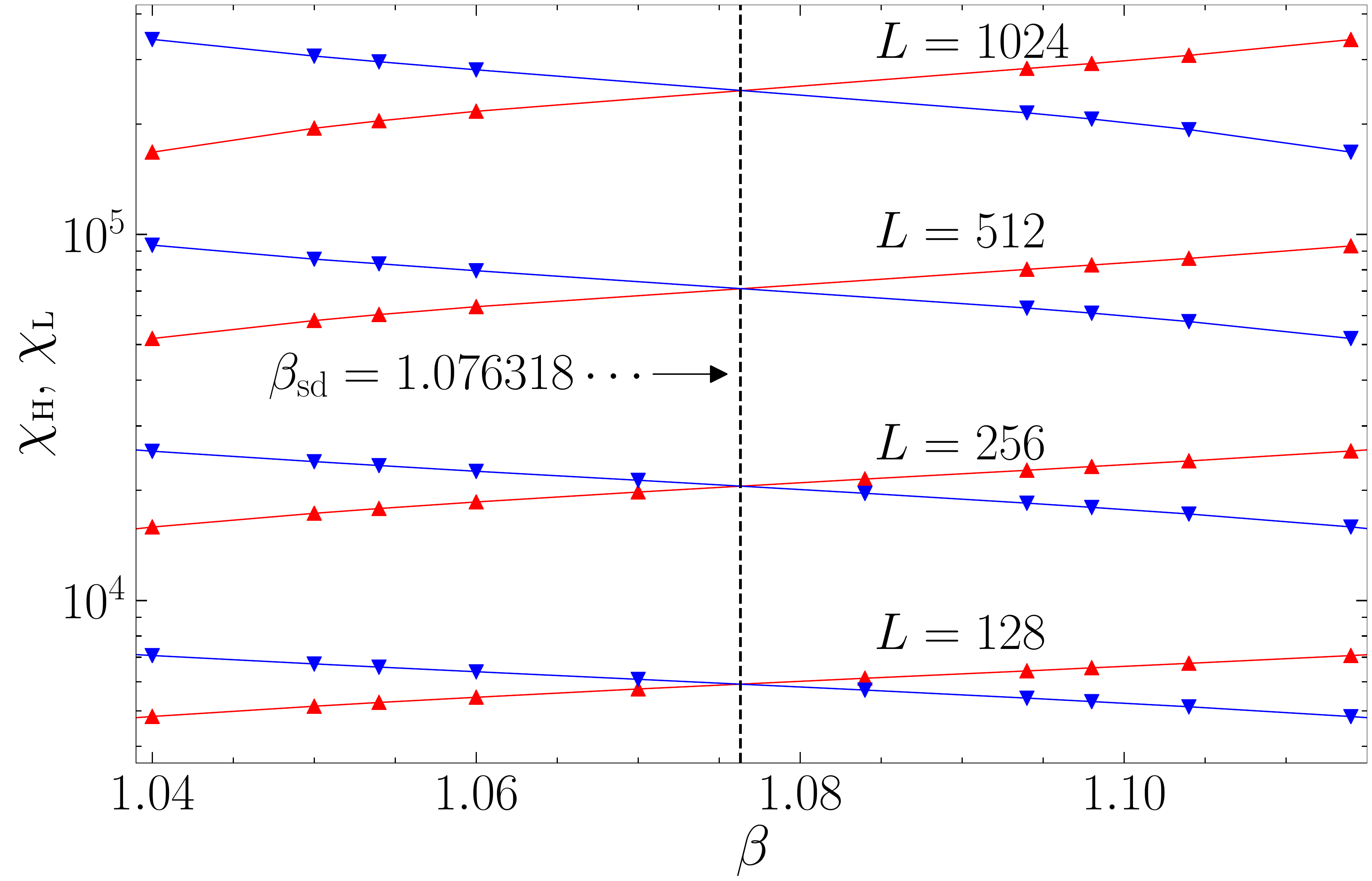}
    \caption{Linear plot of $\chi_\textsc{h}$ and $\chi_\textsc{l}$ versus $\beta$ for $q=5$. The red dots and blue dots correspond to $\chi_\textsc{h}$ and $\chi_\textsc{l}$, respectively, with system size $L= 128, 256, 512,1024$ (from bottom to top). The black vertical dashed line shows the exact self-dual point $\beta_{\rm sd} = 1.076\,318\cdots$ .}
    \label{fig: chiHL_q5}
\end{figure}
In Secs.~\ref{Subsec: critical_points} and \ref{subsec:critical_points_duality} we found that there is still a duality interconnecting the critical points $\bco$ and $\bct$ for $q\geq 5$ despite the fact that the model is no longer strictly self-dual. 
Now we would like to extend the definition of the self-dual point $\beta_{\rm sd}$ and find an approximate one for $q > 5$. 
Here, we define it to be the point at which $\chi_\textsc{h}(\beta, L)$ and $\chi_\textsc{l}(\beta, L)$ are identical, i.e., 
$\chi_{\rm diff}(\beta_{\rm sd}, L)\equiv \chi_\textsc{h}(\beta_{\rm sd}, L) - \chi_\textsc{l}(\beta_{\rm sd}, L) = 0$.
The self-dual point in the thermodynamic limit is then obtained by $\beta_{\rm sd} = \lim_{L\to \infty}\beta_{\rm sd}(L)$.
Notice that we, in principle, only require $\chih$ and $\chil$ to have the same scaling behavior at $\beta_{\rm sd}(L)$ for $q > 5$. Our definition is more stringent by demanding that their amplitudes should also be equal at the self-dual point.
This stringent definition recovers the exact self-dual point for $q = 5$:
As shown in Fig.~\ref{fig: chiHL_q5}, the intersections of $\chi_\textsc{h}(\beta, L)$ and $\chi_\textsc{l}(\beta, L)$ for $q=5$
are almost independent of the system size and consistent with the exact self-dual point $\beta_{\rm sd}= 1.076\,318\dots$.
\begin{figure}[t]
	\centering
	\includegraphics[width=.95\columnwidth]{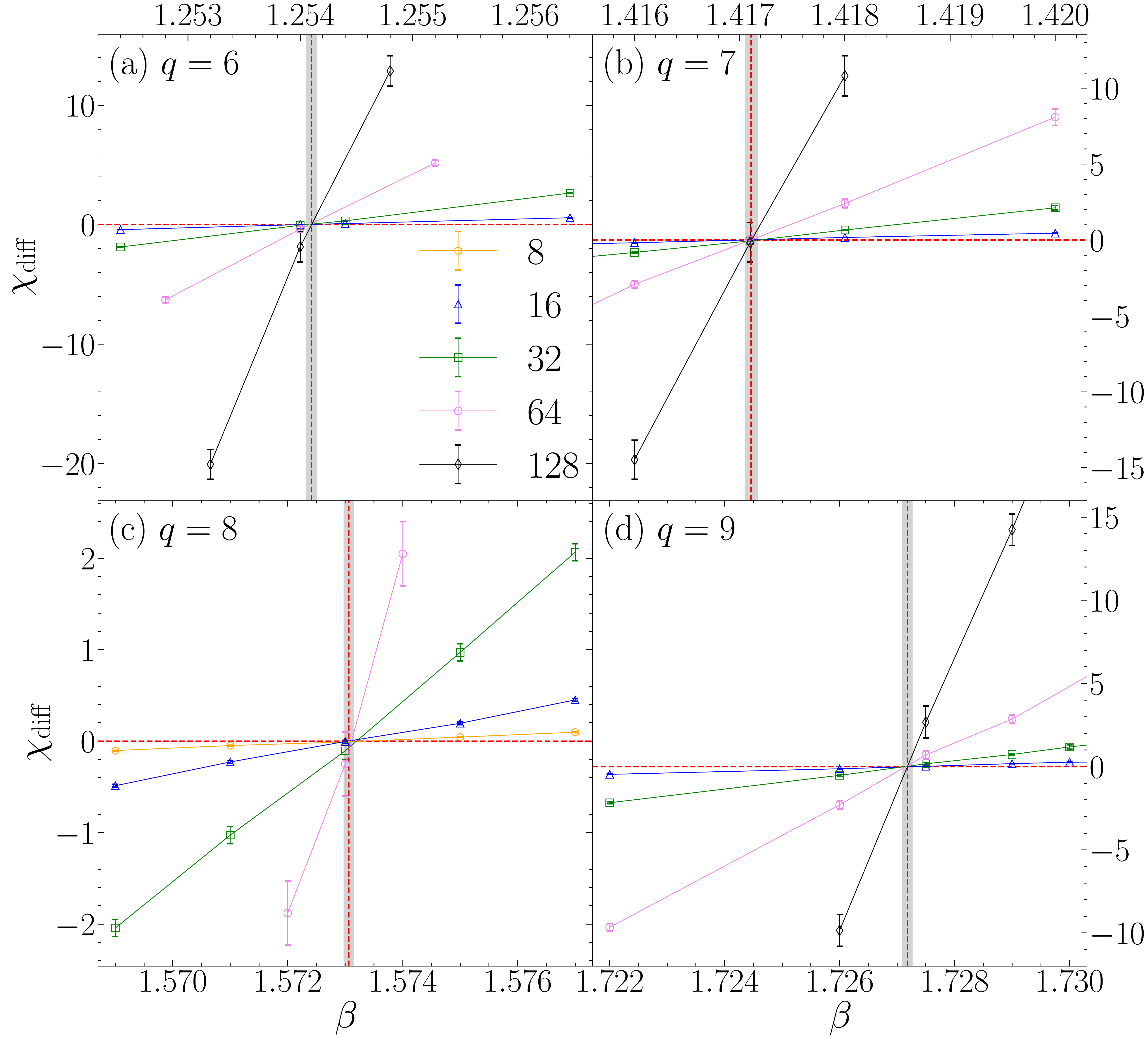}
	\caption{Plots of $\chi_{\rm diff}$ versus the inverse temperature $\beta$ for (a) $q=6$, (b) $q=7$, (c) $q=8$, and (d) $q=9$. The vertical red line represents the central value of $\beta_{\rm sd}$ and the shadow shows the error bar.}
	\label{fig:chiD}
\end{figure}
For $q = 6$--9, we plot $\chi_{\rm diff}(\beta, L)$ for several system sizes in Fig.~\ref{fig:chiD}. 
For all values of $q$, $\chi_{\rm diff}$ has excellent intersections even for small system sizes,
which suggests the fitting \textit{Ansatz}
\begin{equation}
	\chi_{\rm diff}  = a_0 + a_1 (\beta-\beta_{\rm sd})L^{ y_d} ,
	% + b_1 L^{y_1}\\
	% &+ c_1 (K-\beta_{\rm sd})L^{2y_t} +d_1 L^{y_1+y_t}(K-\beta_{\rm sd}).
	\label{eq:fitting_ansatz_beta_sd}
\end{equation}
where $a_0$ is a constant and should be consistent with $0$. The fitting results are summarized in Table~\ref{tab: Ksd}. 
In the fits of $\chi_{\rm diff}$, we first leave $a_0$ and $a_1$ free and find that all the estimates of $a_0$ are consistent with 0. 
Then we set $a_0 = 0$ and get consistent estimates of $\beta_{\rm sd}$. Note that no finite-size corrections are included in Eq.~\eqref{eq:fitting_ansatz_beta_sd}, implying that no shifting of approximate self-dual points $\beta_{\rm sd}(L)$ is observed for different linear system sizes $L$.

\begin{table}[h]
\centering
\caption{Fitting results of the self-dual point $\beta_{\rm sd}$ from $\chi_{\rm diff} = \chi_\textsc{h} - \chi_\textsc{l}$ for $q=6,7,8,9$ with the \textit{Ansatz}~\eqref{eq:fitting_ansatz_beta_sd}.}
\scalebox{0.96}{
\begin{tabular}{|l|lllllc|}
\hline
\multicolumn{1}{|l|}{$q$}   & \multicolumn{1}{c}{$L_{\rm m}$} & \multicolumn{1}{c}{$\beta_{\rm sd}$}  & \multicolumn{1}{c}{$y_d$} & \multicolumn{1}{c}{$a_0$} 	& \multicolumn{1}{c}{$a_1$} & \multicolumn{1}{c|}{${\chi}^2/{\mathcal{D}}$}\\
\hline
\multirow{2}{*}{6}
&16    &1.254\,10(1)	     &2.15(1)   	    &~0.015(6)  	&-0.65(3)  	&16.7/17\\
&24    &1.254\,11(2)	     &2.12(2)   	    &~0.02(2)   	&-0.71(5)  	&10.7/13\\
\hline
\multirow{2}{*}{7}
&24   	&1.417\,11(3)	  &2.09(3)  &~-0.00(2)  	&-0.48(6)  	&13.4/16\\
&32   	&1.417\,10(4)	  &2.02(5)  &~-0.02(4)  	&-0.6(1)   	&7.3/12\\
\hline
\multirow{2}{*}{8}
&12  &1.573\,06(4)	  &2.13(3)   	&~-0.001(5) 	&-0.31(3)  	&16.6/22\\
&16  &1.573\,03(5)	  &2.12(4)   	&~-0.01(1)  	&-0.32(4)  	&10.2/17\\
\hline
\multirow{2}{*}{9}
&24  &1.727\,17(4)	  &2.14(2)   	&~-0.02(2)  	&-0.25(2)  	   &26.1/22\\
&32  &1.727\,20(5)	  &2.14(2)   	&~0.01(4)   	&-0.25(2)  	   &18.5/18\\
\hline
\end{tabular}}
\label{tab: Ksd}
\end{table}

We notice that the value of $y_d$ is greater than the lattice dimension $d=2$. This unusual result is actually due to the fact that the
\textit{Ansatz} proposed above does not truly describe the scaling behavior of $\chi_{\rm diff}$. To derive the correct \textit{Ansatz},
let us write $\chih(\beta, L) = L^{2-\eta_\textsc{h}(\beta)}   a_{\textsc{h}}(\beta)$ and $\chil(\beta, L) = L^{2-\eta_{\textsc{l}}(\beta)} a_{\textsc{l}}(\beta)$
in the QLRO phase
and expand them with respect to $\beta$ at $\beta_{\rm sd}$ to first order
\begin{equation}
	\chi(\beta, L) = L^{2-\eta} \left[a - \epsilon \left(a^\prime - a \eta^\prime \ln L \right )\right] 
	+ O(\epsilon^2),
\end{equation}
where $\epsilon = \beta_{\rm sd} - \beta$ and  $(\chi, a, \eta)$ stands for $(\chi_{\textsc{h}}, a_{\textsc{h}}, \eta_{\textsc{h}})$ or $(\chi_{\textsc{l}}, a_{\textsc{l}}, \eta_\textsc{l})$.
All functions are evaluated at $\beta_{\rm sd}$. According to our definition of $\beta_{\rm sd}$, we have $a_{\textsc{h}}(\beta_{\rm sd}) = a_{\textsc{l}}(\beta_{\rm sd})$ and $\eta_{\textsc{h}}(\beta_{\rm sd}) = \eta_{\textsc{l}}(\beta_{\rm sd}) \equiv \eta(\beta_{\rm sd})$. 
Therefore, $\chi_{\rm diff}(\beta, L)$ can be written as
\begin{equation}\label{eq:chi_diff_scaling}
	\chi_{\rm diff}(\beta, L) = a_0 + \epsilon L^{2-\eta(\beta_{\rm sd})}(a_1 \ln L + a_2) + O(\epsilon^2),
\end{equation}
where $a_0$ should be consistent with 0.
From Eq.~\eqref{eq:chi_diff_scaling} we find that the leading scaling term of $\chi_{\rm diff}(\beta, L)$ is actually
$L^{2-\eta(\beta_{\rm sd})} \ln L$, which results in $y_d \approx 2.1$ for the \textit{Ansatz}~\eqref{eq:fitting_ansatz_beta_sd}. We refit the data with Eq.~\eqref{eq:chi_diff_scaling} and only keep the leading scaling term by setting $a_0 = a_2 = 0$. The results are shown in
Table~\ref{tab: Ksd2}. The estimates of $\beta_{\rm sd}$ are consistent with the one obtained via Eq.~\eqref{eq:fitting_ansatz_beta_sd}.
The final estimates of $\beta_{\rm sd}$ are presented in the third column of Table~\ref{tab: summary} and indicated with vertical red lines in Fig.~\ref{fig:chiD}.
We plot our estimated $\beta_{\rm sd}$ as a function of $q$ in Fig.~\ref{fig: SDKc}. The linear fit of the data for $q > 5$ suggests $\beta_{\rm sd}$ scales as $0.31(1)+q/2\pi$ for the large-$q$ limit.

\begin{table}[h]
\centering
\caption{Fitting results of the self-dual point $\beta_{\rm sd}$ from $\chi_{\rm diff} = \chi_\textsc{h} - \chi_\textsc{l}$ for $q=6,7,8,9$ with the \textit{Ansatz}~\eqref{eq:chi_diff_scaling}.}
\begin{tabular}{|l|llllc|}
\hline
\multicolumn{1}{|l|}{$q$}   & \multicolumn{1}{c}{$L_{\rm m}$} & \multicolumn{1}{c}{$\beta_{\rm sd}$}  & \multicolumn{1}{c}{$\eta(\beta_{\rm sd})$} & \multicolumn{1}{c}{$a_1$} & \multicolumn{1}{c|}{${\chi}^2/{\mathcal{D}}$}\\
\hline
\multirow{2}{*}{6}
&16    &1.254078(9) 	&0.15(2)  	&-0.53(3)  &20.1/18\\ 
&24    &1.254087(10)	&0.14(2)   &-0.52(4)  &11.7/14\\ 
\hline
\multirow{2}{*}{7}
&16    &1.41710(2)&0.20(2)  		&-0.38(3)  	&24.7/19\\ 
&24    &1.41711(2)&0.17(3) 		&-0.35(4)  	&12.4/15\\ 
\hline
\multirow{2}{*}{8}
&16    &1.57308(3)	&0.19(4)   &-0.28(3)  	&10.7/18\\ 
&24    &1.57306(4)	&0.19(7)   &-0.28(6)  	&5.5/13\\ 
\hline
\multirow{2}{*}{9}
&24    &1.72721(3)	&0.11(2)   		&-0.17(1)  	&27.2/22\\ 
&32    &1.72719(3) 	&0.11(2)   		&-0.17(2)  	&18.5/18\\ 
\hline
\end{tabular}
\label{tab: Ksd2}
\end{table}

At our estimated self-dual point $\beta_{\rm sd}$, we then obtain the high-accuracy estimate of $\eta(\beta_{\rm sd})$ by applying finite-size analysis to $\chi_\textsc{h}(\beta_{\rm sd}, L)$ [or $\chi_\textsc{l}(\beta_{\rm sd}, L)$]. 
We perform  the least-squares fits to the Monte Carlo data via the \textit{Ansatz}
\begin{equation}
    \chi_\textsc{h}(\beta_{\rm sd}, L) =  L^{2-\eta(\beta_{\rm sd})}(a_0 + d_1 L^{y_1}) + c_0.
    \label{eq:fitting_ansatz_eta_sd}
\end{equation}    
Our results are summarized in Table~\ref{tab:fits_eta_sd} with the final estimates given in Table~\ref{tab: summary}, and we find that the estimated values of $\eta(\beta_{\rm sd})$ agree with $1/q$, as demonstrated in Fig.~\ref{fig: SDKc}(b),
which is consistent with the RG analysis of the $q$-state Villain clock model in Sec.~\ref{sec:Villain}.

\begin{table}[h]
\setlength\tabcolsep{1pt}
\centering
\caption{Fitting results of $\eta(\beta_{\rm sd})$ from  $\chi_\textsc{h}$ for $q=5,6,7,8,9$ with \textit{Ansatz}~\eqref{eq:fitting_ansatz_eta_sd}. 
We finally obtain $\eta(\beta_{\rm sd})= 0.200(2), 0.1665(3), 0.1426(7), 0.1250(2), 0.1111(2)$ for $q=5,6,7,8,9$, respectively. }
\scalebox{0.96}{
% \begin{tabular}{|l|l|cccccc|}
\begin{tabular}{|l|c|llllcc|}
\hline
\multicolumn{1}{|l|}{$q$}   & \multicolumn{1}{c|}{$L_{\rm m}$} & \multicolumn{1}{c}{$\eta(\beta_{\rm sd})$}  & \multicolumn{1}{c}{$a_0$} & \multicolumn{1}{c}{$d_1$} 	& \multicolumn{1}{c}{$c_0$} & \multicolumn{1}{c}{$y_1$}   & \multicolumn{1}{c|}{${\chi}^2/{\mathcal{D}}$}\\
\hline  
\multirow{2}{*}{5}    &8     &~~0.200(1)  	&0.948(6)  	&~0.29(5)   	&~-0.5(1)   	&-0.8(1) 		&1.8/5\\ 
    &12    &~~0.201(1)  	&0.954(7)  	&~0.4(3)    	&~-0.9(6)   	&-1.0(2)   		&1.0/4\\ 
\hline     
\multirow{2}{*}{6}    
    &6     &~~0.166\,4(1) 	&0.9529(7) 	&~0.6(4)    	&~-0.6(4)   	&-1.5(1)   	&7.1/7\\ 
   &8     &~~0.166\,5(2) 	&0.9534(9) 	&~2(7)      	&~-2(8)     	&-1.7(3)   	&6.4/6\\ 
\hline     
\multirow{2}{*}{7}    
    &12    &~~0.143\,1(2) 	&0.9606(7) 	&~0.4(3)    	&\multicolumn{1}{c}{-}		&-2.1(4)   	&3.1/3\\ 
   &16    &~~0.142\,6(7) 	&0.958(4)  	&~0.04(4)   	&\multicolumn{1}{c}{-}		&-1.0(7)   	&1.3/2\\ 
\hline     
\multirow{2}{*}{8}     &6     &~~0.125\,03(6)	&0.9644(2) 	&~-0.4(1)   	&~0.40(8)   	&-2       	&3.8/7\\ 
    &8     &~~0.125\,01(8)	&0.9643(3) 	&~-0.4(2)   	&~0.4(2)    	&-2       	&3.7/6\\ 
\hline     
\multirow{2}{*}{9}    &4     &~~0.111\,12(3)	&0.9679(1) 	&~-0.38(3)  	&~0.39(3)   	&-2        &4.6/9\\ 
    &6     &~~0.111\,14(4)	&0.9680(2) 	&~-0.33(10) 	&~0.35(8)   	&-2        &4.3/8\\  
\hline 
\end{tabular}}
\label{tab:fits_eta_sd} 
\end{table}

\section{Conclusion}
\label{Sec:Conclusion}
In this work, we applied the worm algorithm to simulate the $q$-state clock model with $5\le q \le 9$ in its two flow representations obtained from the
high-$T$ and low-$T$ expansions.
By finite-size analysis of the susceptibility-like quantities $\chi_\textsc{h}$ and $\chi_\textsc{l}$, we determined the BKT transition points $\bco$ and $\bct$, which significantly improve the precision of the previous MC estimates.
Near $\beta_{\rm c1}$, the exponential divergence of $\xi$ and the scaling behavior of the magnetic susceptibility $\chi_\textsc{h}$ confirm that the clock model with $q\ge 5$ at $\bco$ is in the universality class of the 2D $XY$ model.
Interestingly, we found that even some nonuniversal parameters are consistent with each other, which implies that the clock models with various $q$ values are adjacent on the critical surface and flow into the same fixed point under the process of renormalization group flows.

Besides the high-precision estimates of the critical points, the flow representation also provided a framework to study the duality of the model both analytically and numerically.
The rich phenomena associated with the duality and self-duality properties of the clock model were vividly illustrated by the following
\begin{enumerate}[(i)]
\item At each BKT transition point, the pair of anomalous dimensions $\eta_1=1/4$ and $\eta_2 = 4/q^2$ can be simultaneously 
observed in the finite-size scaling of susceptibility like quantities for the two flow representations, respectively. More precisely, one has 
$\chi_\textsc{h} (\bco, L) \sim L^{2-\eta_1}$ and $\chi_\textsc{l}(\bco, L) \sim L^{2-\eta_2}$ for the high-$T$ transition point $\bco$ and $\chi_\textsc{h} (\bct, L) \sim L^{2-\eta_2}$ and $\chi_\textsc{l}(\bct, L) \sim L^{2-\eta_1}$ for the low-$T$ transition point $\bct$.
\item For $q=5$--9, we numerically observe the stringent relation $\chih(\bct, L) = \chil(\bco, L)$ for each system size $L$.
\item The stringent condition $\chi_{\rm diff} (\beta_{\rm sd}, L) = 0$ can be used to define the approximate self-dual point $\beta_{\rm sd}(L)$ for $q > 5$, which is found to be nearly $L$-independent
and $\beta_{\rm sd} = \lim_{L\to \infty} \beta_{\rm sd} (L)$ follows $\beta_{\rm sd} \simeq q/2\pi$ in the large-$q$ limit.
\item At the approximate self-dual point $\beta_{\rm sd}$, we found that the anomalous dimension $\eta(\beta_{\rm sd})$ agrees with $1/q$, which is the same as that of the Villain clock model at its exact self-dual point. On this basis, we conjectured that $\eta(\beta_{\rm sd})=1/q$ holds exactly at our defined self-dual point and is universal for systems in the $q$-state clock universality class.
\end{enumerate}

Actually, for each point in the QLRO phase $\beta \in [\bco, \bct]$, a pair of anomalous dimensions ($\eta$ and  $\eta^*$) can be observed for the high-$T$ and low-$T$ flows, respectively. Further, a dual temperature $\beta^*(\beta)$ also exists such that the pair of exponents ($\eta$ and $\eta^*$) also occurs, but for the low-T and high-$T$ flows, respectively. 
These findings enrich our understanding of the critical phenomena and duality of the clock model. 

\begin{table*}[ht]
\centering
\caption{Fitting results of the critical points $\bco$ and $\bct$ from $\chi_\textsc{h}$ and $\chil$ with the \textit{Ansatz}~\eqref{eq:fitting_ansatz_beta_c}.
The column of $\beta_c$ corresponds to $\bco$ when $\mathcal{O}$ is taken as $\chih$ and corresponds to $\bct$ when $\mathcal{O}$ is taken as $\chil$.}
\begin{tabular}{|l|l|llccccccccc|}
\hline
$\mathcal{O}$ &$q$ & $L_{\rm m}$  	&\multicolumn{1}{c}{$\beta_c$} 	&$C_1$ 	&$C_2$ 	&$a_0$ 	&$a_1$ 	&$a_2$ 	&$a_3$ & $n_0$ & $d_1$ &${\chi}^2/{\mathcal{D}}$ 	\\
\hline
\multirow{8}{*}{$\chih$}
&\multirow{4}{*}{6} 
&48  &1.109\,2(3) 	&4.8(2)    	&1.3(2)    	&0.807(3)  	&-0.067(3) 	&-0.0071(8)	&-0.0013(3)	&-0.02(7)&-  	&26.3/22\\ 
&&64  &1.109\,3(4) 	&4.6(3)    	&1.4(3)    	&0.809(4)  	&-0.066(3) 	&-0.007(1) 	&-0.0012(3)	&0.0(1)  &-  	&24.3/17\\ 
&&32  &1.110\,3(5) 	&3.5(5)    	&1.38(5)   	&0.824(7)  	&-0.066(1) 	&-0.0066(3)	&-0.0012(2)	 &- 	&0.08(3)   	   &27.4/27\\
&&48  &1.110\,5(8) 	&3.3(8)    	&1.42(8)   	&0.83(1)   	&-0.065(2) 	&-0.0064(4)	&-0.0012(2)	 &- 	&0.09(6)   	    &24.0/22\\
\cline{2-13}

&\multirow{4}{*}{9} 
&32    &1.120\,2(7) 	&3.4(5)    	&2.1(3)    	&0.823(8)  	&-0.053(3) 	&-0.0039(7)	&-0.0004(2)	&0.24(10)	&0.07(2) &24.2/24\\ 
&&48    &1.120(1)  	&3.7(9)    	&1.7(4)    	&0.82(1)   	&-0.056(3) 	&-0.005(1) 	&-0.0005(3)	&0.1(1)    	&0.07(3) &20.5/19\\ 
&&48    &1.118\,8(3) 	&4.6(2)    	&1.31(4)   	&0.807(3)  	&-0.0605(7)	&-0.0062(4)	&-0.0008(2)	 &- &- &22.1/21\\ 
&&64    &1.118\,9(4) 	&4.5(3)    	&1.34(5)   	&0.808(4)  	&-0.0600(9)	&-0.0063(4)	&-0.0009(2)	 &- &- &17.2/16\\ 
\hline

\multirow{10}{*}{$\chil$}
&\multirow{3}{*}{6}
&48    &1.427\,6(4) 	&4.9(3)    	&1.40(4)   	&0.802(3)  	&0.0475(6) 	&-0.0025(1)	&-&-&-	&23.2/24\\ 
&&64    &1.427\,8(5) 	&5.0(4)    	&1.39(5)   	&0.801(5)  	&0.0475(7) 	&-0.0025(1)	&-&-&-	&19.2/19\\ 
\cline{2-13}
&\multirow{2}{*}{7}
&48    &1.851\,1(6) 	&4.9(2)    	&1.59(3)   	&0.796(3)  	&0.0289(3) 	&-0.00103(4) &-&-&-	&24.8/18\\ 
&&64    &1.850\,9(8) 	&4.7(3)    	&1.55(5)   	&0.797(4)  	&0.0292(4) 	&-0.00105(5) &-&-&-	&18.6/14\\ 
\cline{2-13}
&\multirow{2}{*}{8} 
&32    			&2.349\,8(7) 	&5.0(2)    	&1.61(3)   	&0.791(2)  	&0.0206(2) 	&-0.00048(2) &-&-&-		&43.2/33\\ 
&&48          	&2.348(1)  	&4.5(3)    	&1.69(4)   	&0.797(4)  	&0.0203(2) 	&-0.00048(2) &-&-&-		&34.0/27\\ 

\cline{2-13}
&\multirow{2}{*}{9} 
&24    &2.919\,6(7) 	&5.2(1)    	&1.59(4)   	&0.785(2)  	&0.0159(2) 	&-0.00010(9)	&-&-&-	&24.0/18\\ 
&&32    &2.919\,6(10)	&5.3(2)    	&1.60(5)   	&0.785(3)  	&0.0158(3) 	&-0.00008(9)	&-&-&-	&23.2/15\\ 

\hline
\end{tabular}
\label{tab: chiHL_appendix}
\end{table*}
\begin{acknowledgments}
Y. D. acknowledges support from the National Natural Science Foundation of China (under Grant No. 11625522), the Science and Technology Committee of Shanghai (under Grant No. 20DZ2210100), and the National Key R\&D Program of
China (under Grant No. 2018YFA0306501). 
\end{acknowledgments}

%========================================================================$
%  Appendix
%========================================================================$
\appendix 
\section{The fitting details of $\beta_{c1}$ and $\beta_{c2}$}
\label{Sec: Fitting_details}
In this appendix, we describe the fitting details for estimating the higher transition point $\beta_{\rm c1}$ for $q=6,9$ from $\chi_\textsc{h}$ and the lower transition point  $\beta_{\rm c2}$ for $q=6,7,8,9$ from $\chi_\textsc{l}$. 
For $\beta_{\rm c1}$, we first plot the scaled magnetic susceptibility $\tilde{\chi}_{\textsc{h}} = \chih(\beta, L) L^{-7/4}(\ln L+C_1)^{-1/8}$ versus $\beta$ for $q=6,9$. 

\begin{figure}[tb]
\centering
\includegraphics[width=0.95\columnwidth]{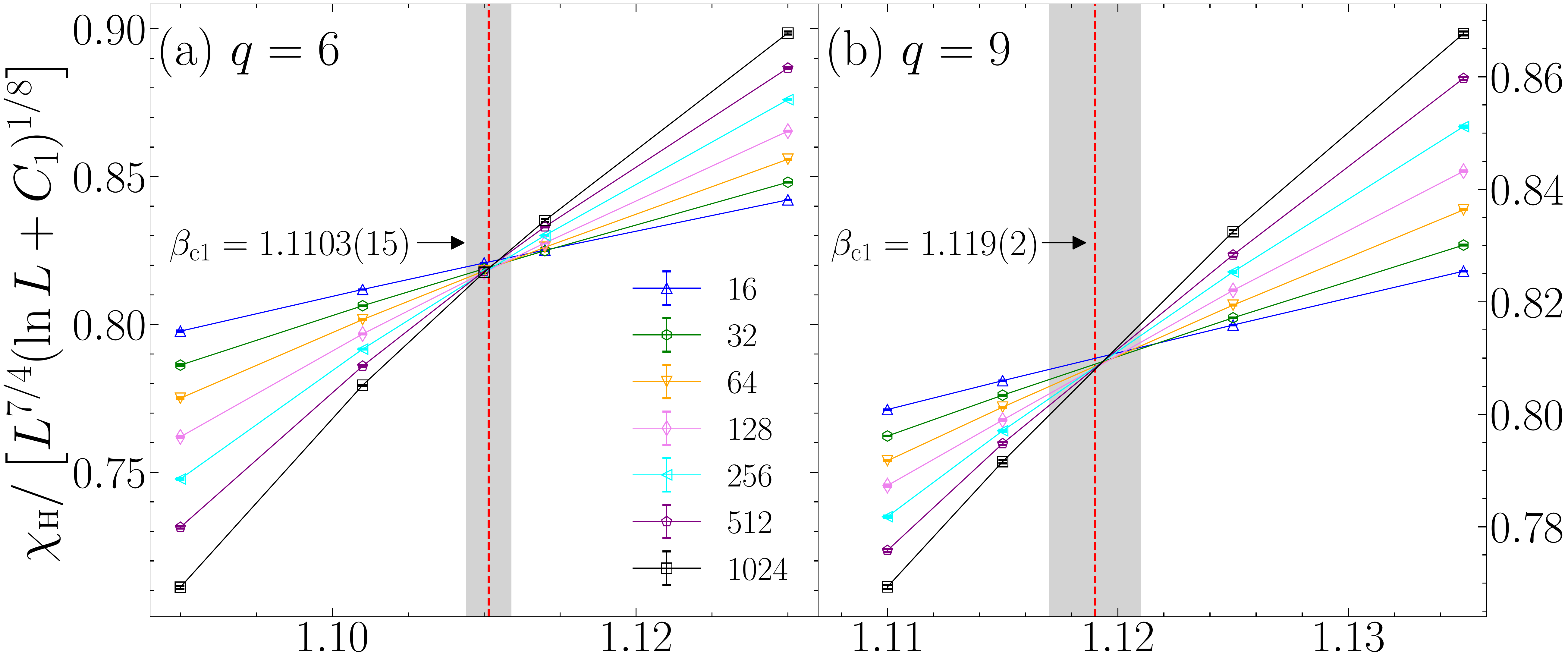}
\caption{Linear plot of the scaled magnetic susceptibility $\chi_\textsc{h}(\bco, L) L^{-7/4}(\ln L+C_1)^{-1/8}$ versus the inverse temperature $\beta$ for (a) $q=6$ and (b) $q=9$.
The constant $C_1$ is set to the corresponding estimated value for each case. The vertical red line represents the central value of $\bco$, and the shadow shows the error bar.}
\vspace{-3pt}
\label{fig: chiH}
\end{figure}

As Fig.~\ref{fig: chiH} shows, both of them have excellent intersections, which implies that $\chih$ suffers from weak finite-size corrections. For $q=6$, we perform the least-squares fit of the MC data via 
the \textit{Ansatz}~\eqref{eq:fitting_ansatz_beta_c}. Similar to $q=5$, we first leave $\beta_{\rm c}$, $C_1$, $C_2$, etc., free and 
find that the effect of the parameters $n_0$, $n_1$, $d_1$ and $d_2$ is negligible, i.e., their amplitudes are consistent with 0.
% their error bars are larger than their center values.
We then only include a single additive correction term by setting $n_1=d_1=d_2=0$ and leaving $n_0$ free, which gives the estimate $\beta_{\rm c1}= 1.1093(4)$. 
Consistent estimates are also obtained by setting $n_0=n_1=d_2=0$ and leaving $d_1$ free, and we find that the value of $d_1$ is nearly consistent with 0. By comparing estimates from various \textit{Ans{\"a}tze}, we conclude that $\beta_{\rm c1} = 1.1103(15)$. 
Similar analysis has been applied to $q=9$, and we get the final estimate $\beta_{\rm c1} = 1.119(2)$. 
These results are reported in Table~\ref{tab: chiHL_appendix}. 
In Table~\ref{tab: chiHL_appendix}, we find that the nonuniversal parameters $C_1$, $C_2$ and $a_0$ for $q=6$ and 9 are numerically consistent,
and the high-$T$ transition point $\beta_{\rm c1}$ quickly approaches the 2D $XY$ model's BKT point $\beta_{\rm BKT}=1.119\,96(6)$ \cite{komuraLargescaleMonteCarlo2012} as $q$ increases.

We then determine the values of $\beta_{\rm c2}$ for $q=6,7,8,9$. Analogous to the case of estimating $\bco$ for $q=6$ and 9, we find that $\chi_{\textsc{l}}$ suffers from weak finite-size corrections compared to $q=5$. 
The fitting results of $\beta_{\rm c2}$ are summarized in Table~\ref{tab: chiHL_appendix}. After considering the systematic errors from using different \textit{Ans{\"a}tze}, we get the final estimates 
$\beta_{\rm c2}=1.4275(8)$, 1.851(1), 2.349(2), 2.920(2) for $q=6,7,8,9$, respectively. 
The same phenomenon that nonuniversal parameters $C_1, C_2$, and $a_0$ are consistent for different values of $q$ is also observed for $\chil$.
Moreover, it is surprising that these nonuniversal parameters of $\chi_\textsc{l}$ numerically agree with those of $\chi_\textsc{h}$, which further demonstrates the duality between the two critical points. 
\bibliography{bibliography.bib}

\begin{thebibliography}{45}%
\makeatletter
\providecommand \@ifxundefined [1]{%
 \@ifx{#1\undefined}
}%
\providecommand \@ifnum [1]{%
 \ifnum #1\expandafter \@firstoftwo
 \else \expandafter \@secondoftwo
 \fi
}%
\providecommand \@ifx [1]{%
 \ifx #1\expandafter \@firstoftwo
 \else \expandafter \@secondoftwo
 \fi
}%
\providecommand \natexlab [1]{#1}%
\providecommand \enquote  [1]{``#1''}%
\providecommand \bibnamefont  [1]{#1}%
\providecommand \bibfnamefont [1]{#1}%
\providecommand \citenamefont [1]{#1}%
\providecommand \href@noop [0]{\@secondoftwo}%
\providecommand \href [0]{\begingroup \@sanitize@url \@href}%
\providecommand \@href[1]{\@@startlink{#1}\@@href}%
\providecommand \@@href[1]{\endgroup#1\@@endlink}%
\providecommand \@sanitize@url [0]{\catcode `\\12\catcode `\$12\catcode
  `\&12\catcode `\#12\catcode `\^12\catcode `\_12\catcode `\%12\relax}%
\providecommand \@@startlink[1]{}%
\providecommand \@@endlink[0]{}%
\providecommand \url  [0]{\begingroup\@sanitize@url \@url }%
\providecommand \@url [1]{\endgroup\@href {#1}{\urlprefix }}%
\providecommand \urlprefix  [0]{URL }%
\providecommand \Eprint [0]{\href }%
\providecommand \doibase [0]{https://doi.org/}%
\providecommand \selectlanguage [0]{\@gobble}%
\providecommand \bibinfo  [0]{\@secondoftwo}%
\providecommand \bibfield  [0]{\@secondoftwo}%
\providecommand \translation [1]{[#1]}%
\providecommand \BibitemOpen [0]{}%
\providecommand \bibitemStop [0]{}%
\providecommand \bibitemNoStop [0]{.\EOS\space}%
\providecommand \EOS [0]{\spacefactor3000\relax}%
\providecommand \BibitemShut  [1]{\csname bibitem#1\endcsname}%
\let\auto@bib@innerbib\@empty
%</preamble>
\bibitem [{\citenamefont {Wu}(1982)}]{wuPottsModel1982}%
  \BibitemOpen
  \bibfield  {author} {\bibinfo {author} {\bibfnamefont {F.~Y.}\ \bibnamefont
  {Wu}},\ }\bibfield  {title} {\bibinfo {title} {The {{Potts}} model},\ }\href
  {https://doi.org/10.1103/RevModPhys.54.235} {\bibfield  {journal} {\bibinfo
  {journal} {Rev. Mod. Phys.}\ }\textbf {\bibinfo {volume} {54}},\ \bibinfo
  {pages} {235} (\bibinfo {year} {1982})}\BibitemShut {NoStop}%
\bibitem [{\citenamefont {Jos{\'e}}\ \emph {et~al.}(1977)\citenamefont
  {Jos{\'e}}, \citenamefont {Kadanoff}, \citenamefont {Kirkpatrick},\ and\
  \citenamefont {Nelson}}]{joseRenormalizationVorticesSymmetrybreaking1977}%
  \BibitemOpen
  \bibfield  {author} {\bibinfo {author} {\bibfnamefont {J.~V.}\ \bibnamefont
  {Jos{\'e}}}, \bibinfo {author} {\bibfnamefont {L.~P.}\ \bibnamefont
  {Kadanoff}}, \bibinfo {author} {\bibfnamefont {S.}~\bibnamefont
  {Kirkpatrick}},\ and\ \bibinfo {author} {\bibfnamefont {D.~R.}\ \bibnamefont
  {Nelson}},\ }\bibfield  {title} {\bibinfo {title} {Renormalization, vortices,
  and symmetry-breaking perturbations in the two-dimensional planar model},\
  }\href {https://doi.org/10.1103/PhysRevB.16.1217} {\bibfield  {journal}
  {\bibinfo  {journal} {Phys. Rev. B}\ }\textbf {\bibinfo {volume} {16}},\
  \bibinfo {pages} {1217} (\bibinfo {year} {1977})}\BibitemShut {NoStop}%
\bibitem [{\citenamefont {Elitzur}\ \emph {et~al.}(1979)\citenamefont
  {Elitzur}, \citenamefont {Pearson},\ and\ \citenamefont
  {Shigemitsu}}]{elitzurPhaseStructureDiscrete1979}%
  \BibitemOpen
  \bibfield  {author} {\bibinfo {author} {\bibfnamefont {S.}~\bibnamefont
  {Elitzur}}, \bibinfo {author} {\bibfnamefont {R.~B.}\ \bibnamefont
  {Pearson}},\ and\ \bibinfo {author} {\bibfnamefont {J.}~\bibnamefont
  {Shigemitsu}},\ }\bibfield  {title} {\bibinfo {title} {Phase structure of
  discrete {{Abelian}} spin and gauge systems},\ }\href
  {https://doi.org/10.1103/PhysRevD.19.3698} {\bibfield  {journal} {\bibinfo
  {journal} {Phys. Rev. D}\ }\textbf {\bibinfo {volume} {19}},\ \bibinfo
  {pages} {3698} (\bibinfo {year} {1979})}\BibitemShut {NoStop}%
\bibitem [{\citenamefont {Cardy}(1980)}]{Cardy1980General}%
  \BibitemOpen
  \bibfield  {author} {\bibinfo {author} {\bibfnamefont {J.~L.}\ \bibnamefont
  {Cardy}},\ }\bibfield  {title} {\bibinfo {title} {General discrete planar
  models in two dimensions: Duality properties and phase diagrams},\ }\href
  {https://doi.org/10.1088/0305-4470/13/4/037} {\bibfield  {journal} {\bibinfo
  {journal} {J. Phys. A Math.}\ }\textbf {\bibinfo {volume} {13}},\ \bibinfo
  {pages} {1507} (\bibinfo {year} {1980})}\BibitemShut {NoStop}%
\bibitem [{\citenamefont {Brito}\ \emph {et~al.}(2010)\citenamefont {Brito},
  \citenamefont {Redinz},\ and\ \citenamefont
  {Plascak}}]{britoTwodimensionalClockModels2010}%
  \BibitemOpen
  \bibfield  {author} {\bibinfo {author} {\bibfnamefont {A.~F.}\ \bibnamefont
  {Brito}}, \bibinfo {author} {\bibfnamefont {J.~A.}\ \bibnamefont {Redinz}},\
  and\ \bibinfo {author} {\bibfnamefont {J.~A.}\ \bibnamefont {Plascak}},\
  }\bibfield  {title} {\bibinfo {title} {Two-dimensional {{XY}} and clock
  models studied via the dynamics generated by rough surfaces},\ }\href
  {https://doi.org/10.1103/PhysRevE.81.031130} {\bibfield  {journal} {\bibinfo
  {journal} {Phys. Rev. E}\ }\textbf {\bibinfo {volume} {81}},\ \bibinfo
  {pages} {031130} (\bibinfo {year} {2010})}\BibitemShut {NoStop}%
\bibitem [{\citenamefont
  {Tobochnik}(1982)}]{tobochnikPropertiesStateClock1982}%
  \BibitemOpen
  \bibfield  {author} {\bibinfo {author} {\bibfnamefont {J.}~\bibnamefont
  {Tobochnik}},\ }\bibfield  {title} {\bibinfo {title} {Properties of the
  {$q$}-state clock model for {$q$} = 4, 5, and 6},\ }\href
  {https://doi.org/10.1103/PhysRevB.26.6201} {\bibfield  {journal} {\bibinfo
  {journal} {Phys. Rev. B}\ }\textbf {\bibinfo {volume} {26}},\ \bibinfo
  {pages} {6201} (\bibinfo {year} {1982})}\BibitemShut {NoStop}%
\bibitem [{\citenamefont {Challa}\ and\ \citenamefont
  {Landau}(1986)}]{challaCriticalBehaviorSixstate1986}%
  \BibitemOpen
  \bibfield  {author} {\bibinfo {author} {\bibfnamefont {M.~S.~S.}\
  \bibnamefont {Challa}}\ and\ \bibinfo {author} {\bibfnamefont {D.~P.}\
  \bibnamefont {Landau}},\ }\bibfield  {title} {\bibinfo {title} {Critical
  behavior of the six-state clock model in two dimensions},\ }\href
  {https://doi.org/10.1103/PhysRevB.33.437} {\bibfield  {journal} {\bibinfo
  {journal} {Phys. Rev. B}\ }\textbf {\bibinfo {volume} {33}},\ \bibinfo
  {pages} {437} (\bibinfo {year} {1986})}\BibitemShut {NoStop}%
\bibitem [{\citenamefont {Yamagata}\ and\ \citenamefont
  {Ono}(1991)}]{yamagataPhaseTransitions6clock1991}%
  \BibitemOpen
  \bibfield  {author} {\bibinfo {author} {\bibfnamefont {A.}~\bibnamefont
  {Yamagata}}\ and\ \bibinfo {author} {\bibfnamefont {I.}~\bibnamefont {Ono}},\
  }\bibfield  {title} {\bibinfo {title} {Phase transitions of the 6-clock model
  in two dimensions},\ }\href {https://doi.org/10.1088/0305-4470/24/1/033}
  {\bibfield  {journal} {\bibinfo  {journal} {J. Phys. A: Math. Gen.}\ }\textbf
  {\bibinfo {volume} {24}},\ \bibinfo {pages} {265} (\bibinfo {year}
  {1991})}\BibitemShut {NoStop}%
\bibitem [{\citenamefont {Tomita}\ and\ \citenamefont
  {Okabe}(2002)}]{tomitaProbabilitychangingClusterAlgorithm2002}%
  \BibitemOpen
  \bibfield  {author} {\bibinfo {author} {\bibfnamefont {Y.}~\bibnamefont
  {Tomita}}\ and\ \bibinfo {author} {\bibfnamefont {Y.}~\bibnamefont {Okabe}},\
  }\bibfield  {title} {\bibinfo {title} {Probability-changing cluster algorithm
  for two-dimensional {{XY}} and clock models},\ }\href
  {https://doi.org/10.1103/PhysRevB.65.184405} {\bibfield  {journal} {\bibinfo
  {journal} {Phys. Rev. B}\ }\textbf {\bibinfo {volume} {65}},\ \bibinfo
  {pages} {184405} (\bibinfo {year} {2002})}\BibitemShut {NoStop}%
\bibitem [{\citenamefont {Borisenko}\ \emph {et~al.}(2011)\citenamefont
  {Borisenko}, \citenamefont {Cortese}, \citenamefont {Fiore}, \citenamefont
  {Gravina},\ and\ \citenamefont {Papa}}]{borisenkoNumericalStudyPhase2011}%
  \BibitemOpen
  \bibfield  {author} {\bibinfo {author} {\bibfnamefont {O.}~\bibnamefont
  {Borisenko}}, \bibinfo {author} {\bibfnamefont {G.}~\bibnamefont {Cortese}},
  \bibinfo {author} {\bibfnamefont {R.}~\bibnamefont {Fiore}}, \bibinfo
  {author} {\bibfnamefont {M.}~\bibnamefont {Gravina}},\ and\ \bibinfo {author}
  {\bibfnamefont {A.}~\bibnamefont {Papa}},\ }\bibfield  {title} {\bibinfo
  {title} {Numerical study of the phase transitions in the two-dimensional
  {{$Z$}}(5) vector model},\ }\href
  {https://doi.org/10.1103/PhysRevE.83.041120} {\bibfield  {journal} {\bibinfo
  {journal} {Phys. Rev. E}\ }\textbf {\bibinfo {volume} {83}},\ \bibinfo
  {pages} {041120} (\bibinfo {year} {2011})}\BibitemShut {NoStop}%
\bibitem [{\citenamefont {Krcmar}\ \emph {et~al.}(2020)\citenamefont {Krcmar},
  \citenamefont {Gendiar},\ and\ \citenamefont
  {Nishino}}]{krcmarPhaseTransitionSixstate2016}%
  \BibitemOpen
  \bibfield  {author} {\bibinfo {author} {\bibfnamefont {R.}~\bibnamefont
  {Krcmar}}, \bibinfo {author} {\bibfnamefont {A.}~\bibnamefont {Gendiar}},\
  and\ \bibinfo {author} {\bibfnamefont {T.}~\bibnamefont {Nishino}},\
  }\bibfield  {title} {\bibinfo {title} {Entanglement-entropy study of phase
  transitions in six-state clock model},\ }\href
  {https://doi.org/10.12693/APhysPolA.137.598} {\bibfield  {journal} {\bibinfo
  {journal} {Acta Physica Polonica A}\ }\textbf {\bibinfo {volume} {137}},\
  \bibinfo {pages} {598} (\bibinfo {year} {2020})}\BibitemShut {NoStop}%
\bibitem [{\citenamefont {Li}\ \emph {et~al.}(2022)\citenamefont {Li},
  \citenamefont {Pai},\ and\ \citenamefont
  {Gu}}]{liAccurateSimulationQstate2020}%
  \BibitemOpen
  \bibfield  {author} {\bibinfo {author} {\bibfnamefont {G.}~\bibnamefont
  {Li}}, \bibinfo {author} {\bibfnamefont {K.~H.}\ \bibnamefont {Pai}},\ and\
  \bibinfo {author} {\bibfnamefont {Z.-C.}\ \bibnamefont {Gu}},\ }\bibfield
  {title} {\bibinfo {title} {Tensor-network renormalization approach to the
  $q$-state clock model},\ }\href
  {https://doi.org/10.1103/PhysRevResearch.4.023159} {\bibfield  {journal}
  {\bibinfo  {journal} {Phys. Rev. Research}\ }\textbf {\bibinfo {volume}
  {4}},\ \bibinfo {pages} {023159} (\bibinfo {year} {2022})}\BibitemShut
  {NoStop}%
\bibitem [{\citenamefont {Chen}\ \emph {et~al.}(2017)\citenamefont {Chen},
  \citenamefont {Liao}, \citenamefont {Xie}, \citenamefont {Han}, \citenamefont
  {Huang}, \citenamefont {Cheng}, \citenamefont {Wei}, \citenamefont {Xie},\
  and\ \citenamefont {Xiang}}]{chenPhaseTransitionState2017}%
  \BibitemOpen
  \bibfield  {author} {\bibinfo {author} {\bibfnamefont {J.}~\bibnamefont
  {Chen}}, \bibinfo {author} {\bibfnamefont {H.-J.}\ \bibnamefont {Liao}},
  \bibinfo {author} {\bibfnamefont {H.-D.}\ \bibnamefont {Xie}}, \bibinfo
  {author} {\bibfnamefont {X.-J.}\ \bibnamefont {Han}}, \bibinfo {author}
  {\bibfnamefont {R.-Z.}\ \bibnamefont {Huang}}, \bibinfo {author}
  {\bibfnamefont {S.}~\bibnamefont {Cheng}}, \bibinfo {author} {\bibfnamefont
  {Z.-C.}\ \bibnamefont {Wei}}, \bibinfo {author} {\bibfnamefont {Z.-Y.}\
  \bibnamefont {Xie}},\ and\ \bibinfo {author} {\bibfnamefont {T.}~\bibnamefont
  {Xiang}},\ }\bibfield  {title} {\bibinfo {title} {Phase {{transition}} of the
  {\emph{q}}-{{state clock model}}: {{duality}} and {{tensor
  renormalization}}},\ }\href {https://doi.org/10.1088/0256-307X/34/5/050503}
  {\bibfield  {journal} {\bibinfo  {journal} {Chinese Phys. Lett.}\ }\textbf
  {\bibinfo {volume} {34}},\ \bibinfo {pages} {050503} (\bibinfo {year}
  {2017})}\BibitemShut {NoStop}%
\bibitem [{\citenamefont {Li}\ \emph {et~al.}(2020)\citenamefont {Li},
  \citenamefont {Yang}, \citenamefont {Xie}, \citenamefont {Tu}, \citenamefont
  {Liao},\ and\ \citenamefont
  {Xiang}}]{liCriticalPropertiesTwodimensional2020}%
  \BibitemOpen
  \bibfield  {author} {\bibinfo {author} {\bibfnamefont {Z.-Q.}\ \bibnamefont
  {Li}}, \bibinfo {author} {\bibfnamefont {L.-P.}\ \bibnamefont {Yang}},
  \bibinfo {author} {\bibfnamefont {Z.~Y.}\ \bibnamefont {Xie}}, \bibinfo
  {author} {\bibfnamefont {H.-H.}\ \bibnamefont {Tu}}, \bibinfo {author}
  {\bibfnamefont {H.-J.}\ \bibnamefont {Liao}},\ and\ \bibinfo {author}
  {\bibfnamefont {T.}~\bibnamefont {Xiang}},\ }\bibfield  {title} {\bibinfo
  {title} {Critical properties of the two-dimensional $q$-state clock model},\
  }\href {https://doi.org/10.1103/PhysRevE.101.060105} {\bibfield  {journal}
  {\bibinfo  {journal} {Phys. Rev. E}\ }\textbf {\bibinfo {volume} {101}},\
  \bibinfo {pages} {060105(R)} (\bibinfo {year} {2020})}\BibitemShut {NoStop}%
\bibitem [{\citenamefont {Lapilli}\ \emph {et~al.}(2006)\citenamefont
  {Lapilli}, \citenamefont {Pfeifer},\ and\ \citenamefont
  {Wexler}}]{lapilli2006universality}%
  \BibitemOpen
  \bibfield  {author} {\bibinfo {author} {\bibfnamefont {C.~M.}\ \bibnamefont
  {Lapilli}}, \bibinfo {author} {\bibfnamefont {P.}~\bibnamefont {Pfeifer}},\
  and\ \bibinfo {author} {\bibfnamefont {C.}~\bibnamefont {Wexler}},\
  }\bibfield  {title} {\bibinfo {title} {Universality {{away}} from {{critical
  points}} in {{two-dimensional phase transitions}}},\ }\href
  {https://doi.org/10.1103/PhysRevLett.96.140603} {\bibfield  {journal}
  {\bibinfo  {journal} {Phys. Rev. Lett.}\ }\textbf {\bibinfo {volume} {96}},\
  \bibinfo {pages} {140603} (\bibinfo {year} {2006})}\BibitemShut {NoStop}%
\bibitem [{\citenamefont {Hwang}(2009)}]{hwang2009sixstate}%
  \BibitemOpen
  \bibfield  {author} {\bibinfo {author} {\bibfnamefont {C.-O.}\ \bibnamefont
  {Hwang}},\ }\bibfield  {title} {\bibinfo {title} {Six-state clock model on
  the square lattice: {{Fisher}} zero approach with {{Wang-Landau}} sampling},\
  }\href {https://doi.org/10.1103/PhysRevE.80.042103} {\bibfield  {journal}
  {\bibinfo  {journal} {Phys. Rev. E}\ }\textbf {\bibinfo {volume} {80}},\
  \bibinfo {pages} {042103} (\bibinfo {year} {2009})}\BibitemShut {NoStop}%
\bibitem [{\citenamefont {Baek}\ \emph {et~al.}(2010)\citenamefont {Baek},
  \citenamefont {Minnhagen},\ and\ \citenamefont {Kim}}]{baek2010comment}%
  \BibitemOpen
  \bibfield  {author} {\bibinfo {author} {\bibfnamefont {S.~K.}\ \bibnamefont
  {Baek}}, \bibinfo {author} {\bibfnamefont {P.}~\bibnamefont {Minnhagen}},\
  and\ \bibinfo {author} {\bibfnamefont {B.~J.}\ \bibnamefont {Kim}},\
  }\bibfield  {title} {\bibinfo {title} {Comment on ``{{Six-state}} clock model
  on the square lattice: {{Fisher}} zero approach with {{Wang-Landau}}
  sampling''},\ }\href {https://doi.org/10.1103/PhysRevE.81.063101} {\bibfield
  {journal} {\bibinfo  {journal} {Phys. Rev. E}\ }\textbf {\bibinfo {volume}
  {81}},\ \bibinfo {pages} {063101} (\bibinfo {year} {2010})}\BibitemShut
  {NoStop}%
\bibitem [{\citenamefont {Baek}\ and\ \citenamefont
  {Minnhagen}(2010)}]{baek2010nonkosterlitzthouless}%
  \BibitemOpen
  \bibfield  {author} {\bibinfo {author} {\bibfnamefont {S.~K.}\ \bibnamefont
  {Baek}}\ and\ \bibinfo {author} {\bibfnamefont {P.}~\bibnamefont
  {Minnhagen}},\ }\bibfield  {title} {\bibinfo {title}
  {Non-{{Kosterlitz-Thouless}} transitions for the $q$-state clock models},\
  }\href {https://doi.org/10.1103/PhysRevE.82.031102} {\bibfield  {journal}
  {\bibinfo  {journal} {Phys. Rev. E}\ }\textbf {\bibinfo {volume} {82}},\
  \bibinfo {pages} {031102} (\bibinfo {year} {2010})}\BibitemShut {NoStop}%
\bibitem [{\citenamefont {Borisenko}\ \emph {et~al.}(2012)\citenamefont
  {Borisenko}, \citenamefont {Chelnokov}, \citenamefont {Cortese},
  \citenamefont {Fiore}, \citenamefont {Gravina},\ and\ \citenamefont
  {Papa}}]{borisenko2012phase}%
  \BibitemOpen
  \bibfield  {author} {\bibinfo {author} {\bibfnamefont {O.}~\bibnamefont
  {Borisenko}}, \bibinfo {author} {\bibfnamefont {V.}~\bibnamefont
  {Chelnokov}}, \bibinfo {author} {\bibfnamefont {G.}~\bibnamefont {Cortese}},
  \bibinfo {author} {\bibfnamefont {R.}~\bibnamefont {Fiore}}, \bibinfo
  {author} {\bibfnamefont {M.}~\bibnamefont {Gravina}},\ and\ \bibinfo {author}
  {\bibfnamefont {A.}~\bibnamefont {Papa}},\ }\bibfield  {title} {\bibinfo
  {title} {Phase transitions in two-dimensional {$Z(N)$} vector models for {$N
  > 4$}},\ }\href {https://doi.org/10.1103/PhysRevE.85.021114} {\bibfield
  {journal} {\bibinfo  {journal} {Phys. Rev. E}\ }\textbf {\bibinfo {volume}
  {85}},\ \bibinfo {pages} {021114} (\bibinfo {year} {2012})}\BibitemShut
  {NoStop}%
\bibitem [{\citenamefont {Baek}\ \emph {et~al.}(2013)\citenamefont {Baek},
  \citenamefont {M{\"a}kel{\"a}}, \citenamefont {Minnhagen},\ and\
  \citenamefont {Kim}}]{baek2013residual}%
  \BibitemOpen
  \bibfield  {author} {\bibinfo {author} {\bibfnamefont {S.~K.}\ \bibnamefont
  {Baek}}, \bibinfo {author} {\bibfnamefont {H.}~\bibnamefont
  {M{\"a}kel{\"a}}}, \bibinfo {author} {\bibfnamefont {P.}~\bibnamefont
  {Minnhagen}},\ and\ \bibinfo {author} {\bibfnamefont {B.~J.}\ \bibnamefont
  {Kim}},\ }\bibfield  {title} {\bibinfo {title} {Residual discrete symmetry of
  the five-state clock model},\ }\href
  {https://doi.org/10.1103/PhysRevE.88.012125} {\bibfield  {journal} {\bibinfo
  {journal} {Phys. Rev. E}\ }\textbf {\bibinfo {volume} {88}},\ \bibinfo
  {pages} {012125} (\bibinfo {year} {2013})}\BibitemShut {NoStop}%
\bibitem [{\citenamefont {Kumano}\ \emph {et~al.}(2013)\citenamefont {Kumano},
  \citenamefont {Hukushima}, \citenamefont {Tomita},\ and\ \citenamefont
  {Oshikawa}}]{kumanoResponseTwistSystems2013}%
  \BibitemOpen
  \bibfield  {author} {\bibinfo {author} {\bibfnamefont {Y.}~\bibnamefont
  {Kumano}}, \bibinfo {author} {\bibfnamefont {K.}~\bibnamefont {Hukushima}},
  \bibinfo {author} {\bibfnamefont {Y.}~\bibnamefont {Tomita}},\ and\ \bibinfo
  {author} {\bibfnamefont {M.}~\bibnamefont {Oshikawa}},\ }\bibfield  {title}
  {\bibinfo {title} {Response to a twist in systems with {$Z_p$} symmetry:
  {{The}} two-dimensional p -state clock model},\ }\href
  {https://doi.org/10.1103/PhysRevB.88.104427} {\bibfield  {journal} {\bibinfo
  {journal} {Phys. Rev. B}\ }\textbf {\bibinfo {volume} {88}},\ \bibinfo
  {pages} {104427} (\bibinfo {year} {2013})}\BibitemShut {NoStop}%
\bibitem [{\citenamefont
  {Chatelain}(2014)}]{chatelainDMRGStudyBerezinskii2014}%
  \BibitemOpen
  \bibfield  {author} {\bibinfo {author} {\bibfnamefont {C.}~\bibnamefont
  {Chatelain}},\ }\bibfield  {title} {\bibinfo {title} {{DMRG} study of the
  {{Berezinskii}}\textendash{{Kosterlitz}}\textendash{{Thouless}} transitions
  of the {{2D}} five-state clock model},\ }\href
  {https://doi.org/10.1088/1742-5468/2014/11/P11022} {\bibfield  {journal}
  {\bibinfo  {journal} {J. Stat. Mech.}\ }\textbf {\bibinfo {volume} {2014}},\
  \bibinfo {pages} {P11022} (\bibinfo {year} {2014})}\BibitemShut {NoStop}%
\bibitem [{\citenamefont {Surungan}\ \emph {et~al.}(2019)\citenamefont
  {Surungan}, \citenamefont {Masuda}, \citenamefont {Komura},\ and\
  \citenamefont {Okabe}}]{surunganBerezinskiiKosterlitzThouless2019}%
  \BibitemOpen
  \bibfield  {author} {\bibinfo {author} {\bibfnamefont {T.}~\bibnamefont
  {Surungan}}, \bibinfo {author} {\bibfnamefont {S.}~\bibnamefont {Masuda}},
  \bibinfo {author} {\bibfnamefont {Y.}~\bibnamefont {Komura}},\ and\ \bibinfo
  {author} {\bibfnamefont {Y.}~\bibnamefont {Okabe}},\ }\bibfield  {title}
  {\bibinfo {title}
  {Berezinskii\textendash{{Kosterlitz}}\textendash{{Thouless}} transition on
  regular and {{Villain}} types of {$q$}-state clock models},\ }\href
  {https://doi.org/10.1088/1751-8121/ab226d} {\bibfield  {journal} {\bibinfo
  {journal} {J. Phys. A: Math. Theor.}\ }\textbf {\bibinfo {volume} {52}},\
  \bibinfo {pages} {275002} (\bibinfo {year} {2019})}\BibitemShut {NoStop}%
\bibitem [{\citenamefont {Hong}\ and\ \citenamefont
  {Kim}(2020)}]{hongLogarithmicFinitesizeScaling2020}%
  \BibitemOpen
  \bibfield  {author} {\bibinfo {author} {\bibfnamefont {S.}~\bibnamefont
  {Hong}}\ and\ \bibinfo {author} {\bibfnamefont {D.-H.}\ \bibnamefont {Kim}},\
  }\bibfield  {title} {\bibinfo {title} {Logarithmic finite-size scaling
  correction to the leading {{Fisher}} zeros in the p -state clock model: {{A}}
  higher-order tensor renormalization group study},\ }\href
  {https://doi.org/10.1103/PhysRevE.101.012124} {\bibfield  {journal} {\bibinfo
   {journal} {Phys. Rev. E}\ }\textbf {\bibinfo {volume} {101}},\ \bibinfo
  {pages} {012124} (\bibinfo {year} {2020})}\BibitemShut {NoStop}%
\bibitem [{\citenamefont {Ortiz}\ \emph {et~al.}(2012)\citenamefont {Ortiz},
  \citenamefont {Cobanera},\ and\ \citenamefont
  {Nussinov}}]{ortizDualitiesPhaseDiagram2012}%
  \BibitemOpen
  \bibfield  {author} {\bibinfo {author} {\bibfnamefont {G.}~\bibnamefont
  {Ortiz}}, \bibinfo {author} {\bibfnamefont {E.}~\bibnamefont {Cobanera}},\
  and\ \bibinfo {author} {\bibfnamefont {Z.}~\bibnamefont {Nussinov}},\
  }\bibfield  {title} {\bibinfo {title} {Dualities and the phase diagram of the
  {$p$}-clock model},\ }\href {https://doi.org/10.1016/j.nuclphysb.2011.09.012}
  {\bibfield  {journal} {\bibinfo  {journal} {Nucl. Phys. B}\ }\textbf
  {\bibinfo {volume} {854}},\ \bibinfo {pages} {780} (\bibinfo {year}
  {2012})}\BibitemShut {NoStop}%
\bibitem [{\citenamefont {Villain}(1975)}]{villain1975theory}%
  \BibitemOpen
  \bibfield  {author} {\bibinfo {author} {\bibfnamefont {J.}~\bibnamefont
  {Villain}},\ }\bibfield  {title} {\bibinfo {title} {Theory of one- and
  two-dimensional magnets with an easy magnetization plane. ii. the planar,
  classical, two-dimensional magnet},\ }\href
  {https://doi.org/10.1051/jphys:01975003606058100} {\bibfield  {journal}
  {\bibinfo  {journal} {J. Phys. France}\ }\textbf {\bibinfo {volume} {36}},\
  \bibinfo {pages} {581} (\bibinfo {year} {1975})}\BibitemShut {NoStop}%
\bibitem [{\citenamefont {{Berezinski{\v{i}}}}(1971)}]{Berezinskii1971}%
  \BibitemOpen
  \bibfield  {author} {\bibinfo {author} {\bibfnamefont {V.~L.}\ \bibnamefont
  {{Berezinski{\v{i}}}}},\ }\bibfield  {title} {\bibinfo {title} {{Destruction
  of Long-range Order in One-dimensional and Two-dimensional Systems having a
  Continuous Symmetry Group I. Classical Systems}},\ }\href@noop {} {\bibfield
  {journal} {\bibinfo  {journal} {Sov. Phys. JETP}\ }\textbf {\bibinfo {volume}
  {32}},\ \bibinfo {pages} {493} (\bibinfo {year} {1971})}\BibitemShut
  {NoStop}%
\bibitem [{\citenamefont {Kosterlitz}\ and\ \citenamefont
  {Thouless}(1973)}]{kosterlitOrderingMetastabilityPhase}%
  \BibitemOpen
  \bibfield  {author} {\bibinfo {author} {\bibfnamefont {J.~M.}\ \bibnamefont
  {Kosterlitz}}\ and\ \bibinfo {author} {\bibfnamefont {D.~J.}\ \bibnamefont
  {Thouless}},\ }\bibfield  {title} {\bibinfo {title} {{Ordering, metastability
  and phase transitions in two-dimensional systems}},\ }\href
  {https://doi.org/10.1088/0022-3719/6/7/010} {\bibfield  {journal} {\bibinfo
  {journal} {J. Phys. C: Solid State Phys.}\ }\textbf {\bibinfo {volume} {6}},\
  \bibinfo {pages} {1181} (\bibinfo {year} {1973})}\BibitemShut {NoStop}%
\bibitem [{\citenamefont
  {Kosterlitz}(1974)}]{kosterlitzCriticalPropertiesTwodimensional1974}%
  \BibitemOpen
  \bibfield  {author} {\bibinfo {author} {\bibfnamefont {J.~M.}\ \bibnamefont
  {Kosterlitz}},\ }\bibfield  {title} {\bibinfo {title} {The critical
  properties of the two-dimensional {{XY}} model},\ }\href
  {https://doi.org/10.1088/0022-3719/7/6/005} {\bibfield  {journal} {\bibinfo
  {journal} {J. Phys. C: Solid State Phys.}\ }\textbf {\bibinfo {volume} {7}},\
  \bibinfo {pages} {1046} (\bibinfo {year} {1974})}\BibitemShut {NoStop}%
\bibitem [{\citenamefont {Chen}\ \emph {et~al.}(2018)\citenamefont {Chen},
  \citenamefont {Xie},\ and\ \citenamefont
  {Yu}}]{chenPhaseTransitionsFivestate2018}%
  \BibitemOpen
  \bibfield  {author} {\bibinfo {author} {\bibfnamefont {Y.}~\bibnamefont
  {Chen}}, \bibinfo {author} {\bibfnamefont {Z.-Y.}\ \bibnamefont {Xie}},\ and\
  \bibinfo {author} {\bibfnamefont {J.-F.}\ \bibnamefont {Yu}},\ }\bibfield
  {title} {\bibinfo {title} {Phase transitions of the five-state clock model on
  the square lattice},\ }\href {https://doi.org/10.1088/1674-1056/27/8/080503}
  {\bibfield  {journal} {\bibinfo  {journal} {Chinese Phys. B}\ }\textbf
  {\bibinfo {volume} {27}},\ \bibinfo {pages} {080503} (\bibinfo {year}
  {2018})}\BibitemShut {NoStop}%
\bibitem [{\citenamefont {Ueda}\ \emph {et~al.}(2020)\citenamefont {Ueda},
  \citenamefont {Okunishi}, \citenamefont {Harada}, \citenamefont
  {Kr\ifmmode~\check{c}\else \v{c}\fi{}m\'ar}, \citenamefont {Gendiar},
  \citenamefont {Yunoki},\ and\ \citenamefont {Nishino}}]{Ueda2020Finite}%
  \BibitemOpen
  \bibfield  {author} {\bibinfo {author} {\bibfnamefont {H.}~\bibnamefont
  {Ueda}}, \bibinfo {author} {\bibfnamefont {K.}~\bibnamefont {Okunishi}},
  \bibinfo {author} {\bibfnamefont {K.}~\bibnamefont {Harada}}, \bibinfo
  {author} {\bibfnamefont {R.}~\bibnamefont {Kr\ifmmode~\check{c}\else
  \v{c}\fi{}m\'ar}}, \bibinfo {author} {\bibfnamefont {A.}~\bibnamefont
  {Gendiar}}, \bibinfo {author} {\bibfnamefont {S.}~\bibnamefont {Yunoki}},\
  and\ \bibinfo {author} {\bibfnamefont {T.}~\bibnamefont {Nishino}},\
  }\bibfield  {title} {\bibinfo {title} {Finite-$m$ scaling analysis of
  berezinskii-kosterlitz-thouless phase transitions and entanglement spectrum
  for the six-state clock model},\ }\href
  {https://doi.org/10.1103/PhysRevE.101.062111} {\bibfield  {journal} {\bibinfo
   {journal} {Phys. Rev. E}\ }\textbf {\bibinfo {volume} {101}},\ \bibinfo
  {pages} {062111} (\bibinfo {year} {2020})}\BibitemShut {NoStop}%
\bibitem [{\citenamefont {Chatterjee}\ \emph {et~al.}(2018)\citenamefont
  {Chatterjee}, \citenamefont {Puri},\ and\ \citenamefont
  {Paul}}]{Chatterjee2018}%
  \BibitemOpen
  \bibfield  {author} {\bibinfo {author} {\bibfnamefont {S.}~\bibnamefont
  {Chatterjee}}, \bibinfo {author} {\bibfnamefont {S.}~\bibnamefont {Puri}},\
  and\ \bibinfo {author} {\bibfnamefont {R.}~\bibnamefont {Paul}},\ }\bibfield
  {title} {\bibinfo {title} {Ordering kinetics in the $q$-state clock model:
  Scaling properties and growth laws},\ }\href
  {https://doi.org/10.1103/PhysRevE.98.032109} {\bibfield  {journal} {\bibinfo
  {journal} {Phys. Rev. E}\ }\textbf {\bibinfo {volume} {98}},\ \bibinfo
  {pages} {032109} (\bibinfo {year} {2018})}\BibitemShut {NoStop}%
\bibitem [{\citenamefont {Prokof'ev}\ \emph {et~al.}(1998)\citenamefont
  {Prokof'ev}, \citenamefont {Svistunov},\ and\ \citenamefont
  {Tupitsyn}}]{prokofevWormAlgorithmQuantum1998}%
  \BibitemOpen
  \bibfield  {author} {\bibinfo {author} {\bibfnamefont {N.}~\bibnamefont
  {Prokof'ev}}, \bibinfo {author} {\bibfnamefont {B.}~\bibnamefont
  {Svistunov}},\ and\ \bibinfo {author} {\bibfnamefont {I.}~\bibnamefont
  {Tupitsyn}},\ }\bibfield  {title} {\bibinfo {title} {``{{Worm}}'' algorithm
  in quantum monte carlo simulations},\ }\href
  {https://doi.org/https://doi.org/10.1016/S0375-9601(97)00957-2} {\bibfield
  {journal} {\bibinfo  {journal} {Physics Letters A}\ }\textbf {\bibinfo
  {volume} {238}},\ \bibinfo {pages} {253} (\bibinfo {year}
  {1998})}\BibitemShut {NoStop}%
\bibitem [{\citenamefont {Prokof'ev}\ and\ \citenamefont
  {Svistunov}(2001)}]{prokofevWormAlgorithmsClassical2001}%
  \BibitemOpen
  \bibfield  {author} {\bibinfo {author} {\bibfnamefont {N.}~\bibnamefont
  {Prokof'ev}}\ and\ \bibinfo {author} {\bibfnamefont {B.}~\bibnamefont
  {Svistunov}},\ }\bibfield  {title} {\bibinfo {title} {Worm {{algorithms}} for
  {{classical statistical models}}},\ }\href
  {https://doi.org/10.1103/PhysRevLett.87.160601} {\bibfield  {journal}
  {\bibinfo  {journal} {Phys. Rev. Lett.}\ }\textbf {\bibinfo {volume} {87}},\
  \bibinfo {pages} {160601} (\bibinfo {year} {2001})}\BibitemShut {NoStop}%
\bibitem [{\citenamefont {Deng}\ \emph {et~al.}(2007)\citenamefont {Deng},
  \citenamefont {Garoni},\ and\ \citenamefont
  {Sokal}}]{dengDynamicCriticalBehavior2007}%
  \BibitemOpen
  \bibfield  {author} {\bibinfo {author} {\bibfnamefont {Y.}~\bibnamefont
  {Deng}}, \bibinfo {author} {\bibfnamefont {T.~M.}\ \bibnamefont {Garoni}},\
  and\ \bibinfo {author} {\bibfnamefont {A.~D.}\ \bibnamefont {Sokal}},\
  }\bibfield  {title} {\bibinfo {title} {Dynamic {{critical behavior}} of the
  {{worm algorithm}} for the {{Ising model}}},\ }\href
  {https://doi.org/10.1103/PhysRevLett.99.110601} {\bibfield  {journal}
  {\bibinfo  {journal} {Phys. Rev. Lett.}\ }\textbf {\bibinfo {volume} {99}},\
  \bibinfo {pages} {110601} (\bibinfo {year} {2007})}\BibitemShut {NoStop}%
\bibitem [{\citenamefont {Hitchcock}\ \emph {et~al.}(2004)\citenamefont
  {Hitchcock}, \citenamefont {S{\o}rensen},\ and\ \citenamefont
  {Alet}}]{hitchcockDualGeometricWorm2004}%
  \BibitemOpen
  \bibfield  {author} {\bibinfo {author} {\bibfnamefont {P.}~\bibnamefont
  {Hitchcock}}, \bibinfo {author} {\bibfnamefont {E.~S.}\ \bibnamefont
  {S{\o}rensen}},\ and\ \bibinfo {author} {\bibfnamefont {F.}~\bibnamefont
  {Alet}},\ }\bibfield  {title} {\bibinfo {title} {Dual geometric worm
  algorithm for two-dimensional discrete classical lattice models},\ }\href
  {https://doi.org/10.1103/PhysRevE.70.016702} {\bibfield  {journal} {\bibinfo
  {journal} {Phys. Rev. E}\ }\textbf {\bibinfo {volume} {70}},\ \bibinfo
  {pages} {016702} (\bibinfo {year} {2004})}\BibitemShut {NoStop}%
\bibitem [{\citenamefont {El{\c c}i}\ \emph {et~al.}(2018)\citenamefont {El{\c
  c}i}, \citenamefont {Grimm}, \citenamefont {Ding}, \citenamefont {Nasrawi},
  \citenamefont {Garoni},\ and\ \citenamefont
  {Deng}}]{elciLiftedWormAlgorithm2018}%
  \BibitemOpen
  \bibfield  {author} {\bibinfo {author} {\bibfnamefont {E.~M.}\ \bibnamefont
  {El{\c c}i}}, \bibinfo {author} {\bibfnamefont {J.}~\bibnamefont {Grimm}},
  \bibinfo {author} {\bibfnamefont {L.}~\bibnamefont {Ding}}, \bibinfo {author}
  {\bibfnamefont {A.}~\bibnamefont {Nasrawi}}, \bibinfo {author} {\bibfnamefont
  {T.~M.}\ \bibnamefont {Garoni}},\ and\ \bibinfo {author} {\bibfnamefont
  {Y.}~\bibnamefont {Deng}},\ }\bibfield  {title} {\bibinfo {title} {Lifted
  worm algorithm for the {{Ising}} model},\ }\href
  {https://doi.org/10.1103/PhysRevE.97.042126} {\bibfield  {journal} {\bibinfo
  {journal} {Phys. Rev. E}\ }\textbf {\bibinfo {volume} {97}},\ \bibinfo
  {pages} {042126} (\bibinfo {year} {2018})}\BibitemShut {NoStop}%
\bibitem [{\citenamefont {Kramers}\ and\ \citenamefont
  {Wannier}(1941)}]{kramersStatisticsTwoDimensionalFerromagnet1941}%
  \BibitemOpen
  \bibfield  {author} {\bibinfo {author} {\bibfnamefont {H.~A.}\ \bibnamefont
  {Kramers}}\ and\ \bibinfo {author} {\bibfnamefont {G.~H.}\ \bibnamefont
  {Wannier}},\ }\bibfield  {title} {\bibinfo {title} {Statistics of the
  {{two-dimensional ferromagnet}}. {{Part I}}},\ }\href
  {https://doi.org/10.1103/PhysRev.60.252} {\bibfield  {journal} {\bibinfo
  {journal} {Phys. Rev.}\ }\textbf {\bibinfo {volume} {60}},\ \bibinfo {pages}
  {252} (\bibinfo {year} {1941})}\BibitemShut {NoStop}%
\bibitem [{\citenamefont {Parisi}(1988)}]{parisi1988statistical}%
  \BibitemOpen
  \bibfield  {author} {\bibinfo {author} {\bibfnamefont {G.}~\bibnamefont
  {Parisi}},\ }\href@noop {} {\emph {\bibinfo {title} {Statistical field
  theory}}},\ Frontiers in Physics\ (\bibinfo  {publisher} {Addison-Wesley,
  Reading, MA},\ \bibinfo {year} {1988})\BibitemShut {NoStop}%
\bibitem [{\citenamefont {Wang}\ \emph {et~al.}(2021)\citenamefont {Wang},
  \citenamefont {Hou}, \citenamefont {Huang},\ and\ \citenamefont
  {Deng}}]{wangPercolationTwodimensionalModel2021}%
  \BibitemOpen
  \bibfield  {author} {\bibinfo {author} {\bibfnamefont {B.-Z.}\ \bibnamefont
  {Wang}}, \bibinfo {author} {\bibfnamefont {P.}~\bibnamefont {Hou}}, \bibinfo
  {author} {\bibfnamefont {C.-J.}\ \bibnamefont {Huang}},\ and\ \bibinfo
  {author} {\bibfnamefont {Y.}~\bibnamefont {Deng}},\ }\bibfield  {title}
  {\bibinfo {title} {Percolation of the two-dimensional {{XY}} model in the
  flow representation},\ }\href {https://doi.org/10.1103/PhysRevE.103.062131}
  {\bibfield  {journal} {\bibinfo  {journal} {Phys. Rev. E}\ }\textbf {\bibinfo
  {volume} {103}},\ \bibinfo {pages} {062131} (\bibinfo {year}
  {2021})}\BibitemShut {NoStop}%
\bibitem [{\citenamefont {Komura}\ and\ \citenamefont
  {Okabe}(2012)}]{komuraLargescaleMonteCarlo2012}%
  \BibitemOpen
  \bibfield  {author} {\bibinfo {author} {\bibfnamefont {Y.}~\bibnamefont
  {Komura}}\ and\ \bibinfo {author} {\bibfnamefont {Y.}~\bibnamefont {Okabe}},\
  }\bibfield  {title} {\bibinfo {title} {Large-scale {{Monte Carlo}} simulation
  of two-dimensional classical {{XY}} model using multiple {{GPUs}}},\ }\href
  {https://doi.org/10.1143/JPSJ.81.113001} {\bibfield  {journal} {\bibinfo
  {journal} {J. Phys. Soc. Jpn.}\ }\textbf {\bibinfo {volume} {81}},\ \bibinfo
  {pages} {113001} (\bibinfo {year} {2012})}\BibitemShut {NoStop}%
\bibitem [{\citenamefont {Harada}\ and\ \citenamefont
  {Kawashima}(1997)}]{Harada1997Universal}%
  \BibitemOpen
  \bibfield  {author} {\bibinfo {author} {\bibfnamefont {K.}~\bibnamefont
  {Harada}}\ and\ \bibinfo {author} {\bibfnamefont {N.}~\bibnamefont
  {Kawashima}},\ }\bibfield  {title} {\bibinfo {title} {Universal jump in the
  helicity modulus of the two-dimensional quantum xy model},\ }\href
  {https://doi.org/10.1103/PhysRevB.55.R11949} {\bibfield  {journal} {\bibinfo
  {journal} {Phys. Rev. B}\ }\textbf {\bibinfo {volume} {55}},\ \bibinfo
  {pages} {R11949} (\bibinfo {year} {1997})}\BibitemShut {NoStop}%
\bibitem [{\citenamefont {Kadanoff}(1978)}]{kadanoffLatticeCoulombGas1978}%
  \BibitemOpen
  \bibfield  {author} {\bibinfo {author} {\bibfnamefont {L.~P.}\ \bibnamefont
  {Kadanoff}},\ }\bibfield  {title} {\bibinfo {title} {Lattice {{Coulomb}} gas
  representations of two-dimensional problems},\ }\href
  {https://doi.org/10.1088/0305-4470/11/7/027} {\bibfield  {journal} {\bibinfo
  {journal} {J. Phys. A: Math. Gen.}\ }\textbf {\bibinfo {volume} {11}},\
  \bibinfo {pages} {1399} (\bibinfo {year} {1978})}\BibitemShut {NoStop}%
\bibitem [{\citenamefont
  {Nienhuis}(1984)}]{nienhuisCriticalBehaviorTwodimensional1984}%
  \BibitemOpen
  \bibfield  {author} {\bibinfo {author} {\bibfnamefont {B.}~\bibnamefont
  {Nienhuis}},\ }\bibfield  {title} {\bibinfo {title} {Critical behavior of
  two-dimensional spin models and charge asymmetry in the {{Coulomb}} gas},\
  }\href {https://doi.org/10.1007/BF01009437} {\bibfield  {journal} {\bibinfo
  {journal} {J. Stat. Phys.}\ }\textbf {\bibinfo {volume} {34}},\ \bibinfo
  {pages} {731} (\bibinfo {year} {1984})}\BibitemShut {NoStop}%
\bibitem [{\citenamefont {Pelissetto}\ and\ \citenamefont
  {Vicari}(2013)}]{pelissettoRenormalizationgroupFlowAsymptotic2013}%
  \BibitemOpen
  \bibfield  {author} {\bibinfo {author} {\bibfnamefont {A.}~\bibnamefont
  {Pelissetto}}\ and\ \bibinfo {author} {\bibfnamefont {E.}~\bibnamefont
  {Vicari}},\ }\bibfield  {title} {\bibinfo {title} {Renormalization-group flow
  and asymptotic behaviors at the {{Berezinskii-Kosterlitz-Thouless}}
  transitions},\ }\href {https://doi.org/10.1103/PhysRevE.87.032105} {\bibfield
   {journal} {\bibinfo  {journal} {Phys. Rev. E}\ }\textbf {\bibinfo {volume}
  {87}},\ \bibinfo {pages} {032105} (\bibinfo {year} {2013})}\BibitemShut
  {NoStop}%
\end{thebibliography}%
\end{document}